\documentclass[superscriptaddress,floatfix,nofootinbib,longbibliography,prx,twocolumn]{revtex4-1}
\usepackage{latexsym}
\usepackage{amssymb}
\usepackage{graphicx}
\usepackage{amsmath}
\usepackage{bm}
\usepackage[colorlinks,
          linkcolor=black,
            citecolor=black,
            urlcolor=blue
           ]{hyperref}
\usepackage{verbatim}
\usepackage{slashed}
\usepackage{mathrsfs}
\usepackage{extarrows}
\usepackage{mathtools}
\usepackage{comment}
\usepackage{mathtools,slashed}

\newcommand{\ed}[1]{\textcolor{blue}{#1}}

\begin{document}

\title{A generalized boundary condition applied to Lieb-Schultz-Mattis type ingappabilities and many-body Chern numbers}

\author{Yuan Yao}
\email{smartyao@issp.u-tokyo.ac.jp}
\affiliation{Institute for Solid State Physics, University of Tokyo, Kashiwa, Chiba 277-8581, Japan}
\author{Masaki Oshikawa}
\affiliation{Institute for Solid State Physics, University of Tokyo, Kashiwa, Chiba 277-8581, Japan}
\affiliation{Kavli Institute for the Physics and Mathematics of the Universe (WPI), University of Tokyo, Kashiwa, Chiba 277-8583, Japan}

\begin{abstract}
We introduce a new boundary condition which renders the flux-insertion argument for the Lieb-Schultz-Mattis type theorems in two or higher dimensions
free from the specific choice of system sizes.
It also enables a formulation of the Lieb-Schultz-Mattis type theorems in arbitrary dimensions in terms of the anomaly in field theories in $1+1$ dimensions
with a bulk correspondence as a BF-theory in $2+1$ dimensions.
Furthermore, we apply the anomaly-based formulation to the constraints on a half-filled spinless fermion on a square lattice with $\pi$ flux, utilizing a time-reversal, magnetic translations and an on-site internal $U(N)$ symmetries.
This demonstrates the role of the time-reversal anomaly on the ingappabilities of a lattice model.
Moreover, by our new boundary condition, we show that the many-body Chern number of this lattice model is non-vanishing as $N\mod2N$ in the presence of $U(N)$ and magnetic translations.
 {
This can be a general mechanism of anomaly-based constraints on quantized Hall conductance, which generally depends on high-energy
physics, from field theory.}
\end{abstract}

\maketitle

\section{Introduction}
Quantifying various phases for quantum many-body systems is a central task in condensed matter and statistical physics. Recent decades have witnessed several significant phase classifications, e.g. topological ordered phases~\cite{Chen:2010aa,Wen:TOreview2013}
and {symmetry-protected topological (SPT) phases} beyond Landau's symmetry-breaking pattern of strongly-correlated systems with non-perturbative interactions~\cite{Gu:2009aa,Pollmann:2012aa}.
Furthermore, symmetries together with filling fractions also constrain low-energy spectrums when critical phases are gapped, which induce the concept of symmetry-protected critical phases~\cite{Furuya:2017aa,Yao:2019aa} that nontrivial critical phases are ingappable with a unique ground state if the symmetries are respected by Hamiltonians.

One of the most important general principles in quantum many-body systems is Lieb-Schultz-Mattis (LSM) theorem~\cite{Lieb:1961aa} and
its generalizations~\cite{Affleck:1986aa,OYA1997,Oshikawa:2000aa, Hastings:2004ab}.
They show the interplay between the global $U(1)_Q$ charge and translation symmetries.
The theorem states, under certain conditions,
an ``ingappability'' of the system, that is, either the presence of
gapless excitations above the ground states or a ground-state degeneracy
in the limit of the large system size.
It is valid for Hamiltonians with appropriate symmetries for
arbitrary strong interactions, and thus is non-perturbative in nature.

Generally in physics, we expect that the bulk property would not depend
on the choices of boundary conditions. If that is the case, we can
use a boundary condition which is convenient for the calculation and
infer physical results which would be valid independent of the boundary
condition.
In many cases, the periodic boundary condition is chosen as a boundary
condition. A typical example is the band theory of electronic structures.

The original proof~\cite{Lieb:1961aa} of the LSM theorem was also based
on the periodic boundary condition.
The LSM theorem in higher dimensions turns out to be more subtle.
While it is easy to see the failure of the original proof in higher
dimensions, the original proof would still work~\cite{Lieb:1961aa,Affleck:1988}
in an ``anisotropic'' thermodynamic limit in which the ratio of the system
sizes in each direction diverges.
However, one might worry that the system is essentially one-dimensional
in such a limit.
An alternative argument based on an adiabatic flux insertion
and gauge invariance, which does not need such an anisotropic limit,
was proposed later~\cite{Oshikawa:2000aa}.
It still depends on several nontrivial assumptions including the
stability of the gap against the flux insertion, and special
choices of system size as we will discuss later.
It was followed by a more rigorous proof~\cite{Hastings:2004ab} of the LSM
theorem in higher dimensions, and its further
mathematical refinements~\cite{NachtergaeleSims}.

Nevertheless, the flux insertion argument is still attractive in
its simplicity, intuitiveness, and connections to other
concepts in physics.
Indeed, some of the recent extensions~\cite{Parameswaran:2013}
of the LSM theorem is based on the flux insertion argument.
Because of this, it would be valuable to improve the flux insertion
approach to the LSM-type theorems.
One of the subtleties in the flux insertion argument was that
the system sizes must meet a special condition: the lengths
in all but one direction must be coprime with the denominator
of the filling fraction.
Although one may consider the ``thermodynamic limit'' with a series
of finite-size systems satisfying this condition keeping the
ratio of length to be of order of $1$, it is desirable to remove
such a rather artificial condition.

In this work, we introduce a new class of boundary conditions, which
we call tilted boundary conditions (TLBC), which is useful for
derivation of the LSM-type theorems in dimensions higher than one.
As we will demonstrate, with TLBC, the flux insertion argument can be
applied without the artificial condition on the system sizes.
This is also the case for the higher symmetry
($SU(N)$) generalizations~\cite{Affleck:1986aa,Yao:2019aa}.

Furthermore, the TLBC reveals previously unnoticed
relations between anomaly in field theory with the LSM-type theorems.
While the LSM theorem has been well understood in the context of
Tomonaga-Luttinger Liquids in one spatial dimension, field-theoretical
understanding of LSM-type theorems in higher dimensions has been rather
limited.
The TLBC allows us to understand the LSM theorem as an anomaly manifestation.
As a futher application of the anomaly-based approach,
we will discuss the implications of the time-reversal anomaly~\cite{Seiberg:2016aa,Witten:2016aa} with a lattice interpretation.
Taking the advantage of the bulk-boundary correspondence between SPT phase and the time-reversal anomaly on its boundary, the anomaly leads to an ingappability constraint
for the half-filled $\pi$-flux system when time-reversal, on-site $U(N)$ and magnetic translational symmetries are respected.
 {Moreover, the TLBC enables us to obtain a constraint on integer quantum Hall conductances. We show that the $\pi$-flux system must have non-vanishing $N\mod2N$ many-body Chern number in the presence of $U(N)$ and magnetic translational symmetries.
Our constraint generalizes those derived for lattice models
known for $N=1$~\cite{Dana:1985aa,AvronYaffe1986,KolRead1993,Lu:2017aa,Matsugatani:2018}.
Furthermore, our proposal can be a general mechanism for non-perturbative restrictions on quantize Hall conductivity from field theory.
It reflects a deep relation between such phenomena with symmetry-broken surface of SPT phases. }

This paper is organized as follows. In Sec.~\ref{u(1)_LSM}, we introduce our new boundary condition TLBC and apply it to the LSM theorem in arbitrary dimensions. In Sec.~\ref{LSMA}, the TLBC is used to generalize the LSM theorem for spin systems. In Sec.~\ref{time_reversal}, we discuss the LSM-type ingappabilities on the half-filled $\pi$-flux square lattice with an onsite $U(N)$ symmetry, magnetic translational symmetries and a time-reversal symmetry by a time-reversal anomaly. As a further application, a constraint on the integer quantum Hall conductances of the gapped $\pi$-flux system in the presence of $U(N)$ symmetry and magnetic translations is obtained by a bulk-boundary correspondence with the geometry of TLBC in Sec.~\ref{IQHE}. In the Appendix, we present a detailed derivation of the constraint.

\section{TLBC and LSM theorem}
\label{u(1)_LSM}

\subsection{Flux insertion with PBC}
\label{sec.flux_PBC}

Let us first review the flux insertion argument~\cite{Oshikawa:2000aa}
for the LSM theorem in $d \geq 2$ dimensions, and some of the problems
in it.
{We are interested in the energy spectrum of a quantum
many-particle system on a periodic lattice.
We assume that the number of particles is exactly conserved, and
consider the the limit of
the large system size (thermodynamic limit) with a fixed
``filling fraction'' (number of perticles per unit cell) $\nu$.

The many-body gap is defined as the energy gap between
the (possibly degenerate, multiple) ground state and
the continuum of excited states, in the thermodynamic limit.
It is not to be confused with the finite-size gap defined by the
gap between the ground state and the lowest excited state in
a finite system of a fixed size.

When the many-body gap vanishes, namely if the continuum of excited
states starts at an infinitesimal energy above the ground state,
the many-particle system is said to be gapless.
On contrary, when the many-body gap is non-vanishing, the system
is said to be gapped.
The LSM theorem is a constraint on the energy spectrum,
in particular the ground-state degeneracy if the system is gapped,
when the filling fraction $\nu$ is not an integer.
}


{For simplicity, let us consider}
the (hyper)cubic lattice in $d$ dimensions with PBC
\begin{eqnarray}
\label{PBC}
{\vec{r}}\sim{\vec{r}+L_i\hat{x}_i},\,\,i=1,2,\cdots,d,
\end{eqnarray}
where $L_i$ is the length along $i$-th unit vector $\hat{x}_i$, and the Hamiltonian is required to possess translational symmetry and $U(1)_Q$ symmetry.
{
We set the filling factor as a rational number $\nu=p/q$,
where $p$ and $q$ are coprime.}
There is an additional charge quantization condition due to the fundamental degrees of freedom being charge one:
\begin{eqnarray}
\label{charge_quantization}
\nu V\in\mathbb{N},
\end{eqnarray}
where $V\equiv\prod_iL_i$ is the total volume, and this condition is also implied by the compactness of global $U(1)_Q$ symmetry.
{
The PBC means that the system has topologically non-trvial loops,
which may be visualized with ``holes'' in a higher-dimensional
embedding.
Each hole can contain a magnetic flux (Aharanov-Bohm flux, AB flux),
which affects the system through the Aharonov-Bohm effect.
Here we will utilize the AB flux $\Phi$ which is enclosed by
the closed loop in $1$-direction.
Such a flux is represented by the vector potential in $1$-direction
$ A_1 = \Phi /L_1$ .

The many-particle system may be gapless or gapped.
If it is gapless, there is nothing more to say in the LSM theorem.
Therefore, we can assume that the system is gapped, namely the
many-body gap separating the continuum of the excited states
from the ground state(s) is non-vanishing (at $\Phi=0$).
Starting from a ground state, which is also a momentum eigenstate,
as the initial state,
we consider an adiabatic insertion of unit flux quantum $\Phi=2\pi$.
Namely, we consider the
application of a time-dependent $U(1)_Q$ gauge field
\[
A_1= \frac{2\pi t}{TL_1},
\]
with $T\rightarrow+\infty$.
After the insertion of the unit flux quantum, the Hamiltonian is
equivalent (under a large gauge transformation) to the original one without
the flux.
Here we make a crucial assumption that the many-body gap
(which was non-vanishing at zero flux, by the initial assumption)
does not vanish during the flux insertion.
It then follows from the adiabaticity
that the final state must be a ground state below the gap.
One might expect the final state after the large gauge transformation
would be identical to the original ground state.}

However, for an incommensurate filling, there is
a nontrivial momentum shift caused by the flux insertion.
For instance, the lattice momentum change along $\hat{x}_1$ is
\begin{eqnarray}
\label{intuitive}
\Delta P^\text{PBC}_1=2\pi \frac{p}{q}L_2L_3\cdots L_d\mod2\pi,
\end{eqnarray}
which is nonzero if $(L_2\cdots L_d)$ is nondivisible by $q$.
In this case, the final state cannot be identical to the initial
ground state.
Restricting the system sizes $L_{2,3,\ldots,d}$ to coprime with $q$,
the statement of the LSM theorem is derived.
Namely, the system is either gapless or has at least $q$ degenerate
ground states below the gap.
However, this artificial restriction on the system size is clearly
undesired. Physically we would expect the same statement to hold
for generic (large) system sizes, even though the above argument
does not lead to a nontrivial restriction if
$L_2L_3\cdots L_d$ is divisible by $q$.

{
Here let us also comment on the nontrivial assumption
that the non-zero many-body gap does not collapse during the flux insersion.
While this assumption, which is crucial for ``flux insertion'' arguments,
has not proven,
it is rather natural, as the change of the hopping terms is only
of the order of $O(1/L_1)$ in the uniform gauge.
In fact, there are some numerical~\cite{Misguich2002} and
analytical~\cite{Watanabe:2018aa} supporting evidences.
Furthermore, some  versions of the LSM theorem, that follow from
the assumption, have been proven
rigorously~\cite{Hastings:2004ab,NachtergaeleSims}.
Although a rigorous justification (or a clarification on the range of
validity) of the assumption is an interesting problem,
it is outside the scope of the present paper.
}

\subsection{Flux insertion with tilted boundary condition}

\begin{figure}
\begin{center}
\includegraphics[width=8.5cm,pagebox=cropbox,clip]{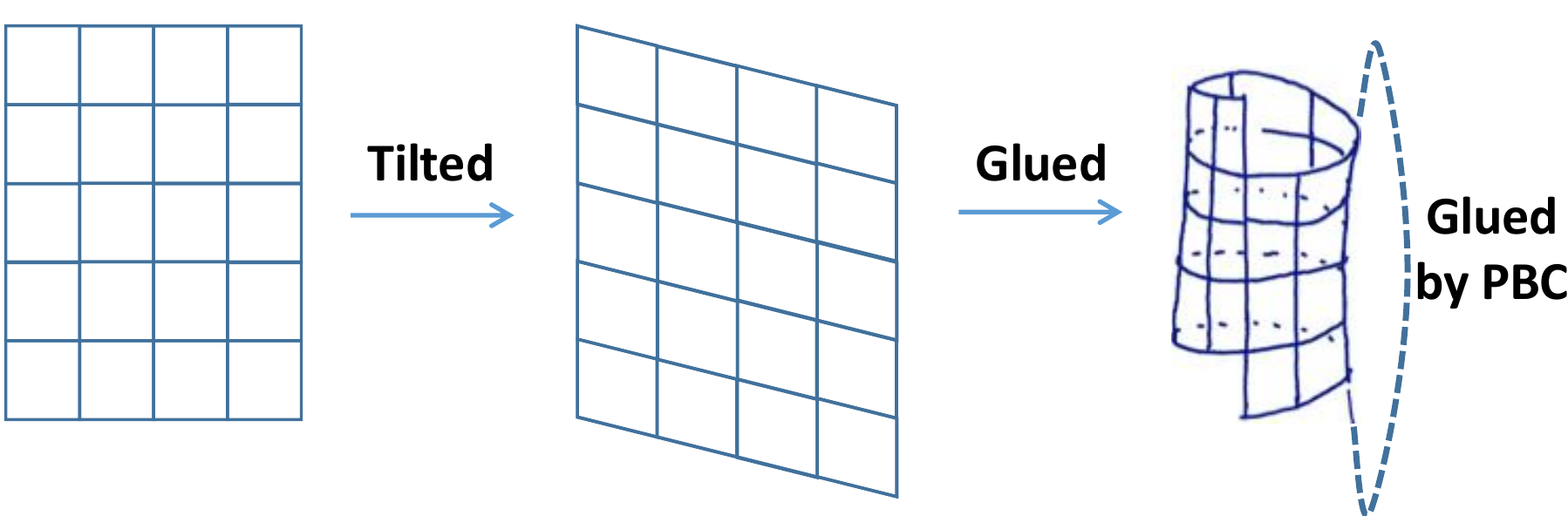}
\caption{TLBC when $d=2$. }
\label{TLBC_torus}
\end{center}
\end{figure}

Here we propose to use, instead of the standard PBC,
the following tilted boundary condition (TLBC)
\begin{eqnarray}
\label{tbc0}
\left\{\begin{array}{ll}\vec{r}+L_i\hat{x}_i\sim\vec{r}+\hat{x}_{i+1},&i=1,\cdots,d-1;\\\vec{r}+L_i\hat{x}_i\sim\vec{r}, &i=d, \end{array}\right.
\end{eqnarray}
where $\hat{x}_i$ is the unit vector along $L_i$.
We sketch the two-dimensional case of TLBC in FIG.~(\ref{TLBC_torus}).
Geometrically, under such identifications, the space is a $d$-dimensional torus.
Combined with the flux insertion and momentum counting, this
leads to the LSM theorem without the artificial restriction on the system
sizes, as we will discuss below.
The TLBC is consistent with all the imposed symmetries---$U(1)_Q$ and translations.
Now let us consider the following flux insertion process $(t\in[0,T])$:
\begin{eqnarray}
\label{flux}
A_i(t)=\frac{2\pi t}{T\prod_{k=i}^dL_k},\,\,i=1,\cdots,d.
\end{eqnarray}
{During the flux insertion,
the gauge~(\ref{flux}) still respects the translational symmetries since it is site-independent.
Thus the canonical momentum remains exactly unchanged during the
time evolution under the flux insertion.

Similarly to the case of PBC,
after completing the flux insertion,
we perform a large gauge transformation
to bring the Hamiltonian back to its original form.
Reflecting the different boundary condition and the different way
of inserting the flux, the large gauge transformation
used for PBC~\cite{Oshikawa:2000aa} does not work as is.
Nevertheless, we find that,
the following large gauge transformation exactly eliminates
the vector potential~\eqref{flux}
at $t=T$, after the flux insertion process:
\begin{eqnarray}
\label{large_U}
\hat{U}_0=\exp\left(\sum_{\vec{r}} i \frac{2\pi r_1\hat{n}_{\vec{r}}}{V}\right),
\end{eqnarray}
where we fix the range of $\vec{r}$ as
\begin{eqnarray}
\label{coordinate_fix}
\vec{r}=r_1\hat{x}_1, \,\,(r_1=1,\cdots,V),
\end{eqnarray}
exhausting all the lattice points.

By applying the large gauge transformation (\ref{large_U}) on the final state, which maps the Hamiltonian back to its original form, we can compare the lattice momenta along $\hat{x}_i$ of the state before and after this insertion process in the same $U(1)_Q$ gauge. The momentum difference is
}
\begin{eqnarray}
\label{LSM}
\Delta P_i&=&\frac{2\pi n}{\prod_{k=i}^dL_k}\mod\left(2\pi\frac{V}{\prod_{k=i}^dL_k}\right)\nonumber\\
&=&{2\pi\nu}\left(\frac{V}{{\prod_{k=i}^{d}L_k}}\right)\mod\left(2\pi\frac{V}{\prod_{k=i}^dL_k}\right),
\end{eqnarray}
where $i=1,2,\cdots,d$, the total number of particles is denoted by $n$ and $n/V=\nu$ is the filling fraction by definition, and the different periods of $\Delta P_i$'s results from the identification of translations as $T_{i+1}=\left(T_i\right)^{L_i}$ due to TLBC.

{
As in the case of the PBC reviewed in Sec.~\ref{sec.flux_PBC},
the momentum shift~\eqref{LSM} implies the LSM
theorem: if $\nu=p/q$ with $p$ and $(q>1)$ coprimes,
the ground states in a gapped phase must be at least $q$-fold degenerate,
as long as the Hamiltonian respects the
$U(1)_Q$ and the lattice translation symmetries.
That is, a trivial insulating phase with a unique ground state is excluded.
We note that, in the present argument, each component of
$\Delta\vec{P}$ gives exactly the same constraint on the ground-state
degeneracies of gapped phases, in contrast to the previous derivation
in Eq.~(\ref{intuitive}).  More importantly, there is no artificial
requirement on the system size beyond the charge quantization condition
of Eq.~(\ref{charge_quantization}).
}

It should be noted, however, the present argument still relies on the
nontrivial assumption that the excitation gap does not collapse
under the adiabatic flux insertion, discussed at the end of
Sec.~\ref{sec.flux_PBC}.

\subsection{Anomaly manifestation of LSM theorem}

The TLBC also reveals a connection between the LSM theorem in
higher dimensions and anomaly in field theory in $1+1$ dimensions.

Let us consider an electronic system, without loss of generality, on a square lattice with TLBC, the lattice point of which can be thought as that of unit cells on a general lattice with translational symmetries.
First, we consider a non-interacting, tight-binding model:
\begin{eqnarray}
\label{T-B}
H_\text{T-B}=-t\sum_{\langle \vec{r},\vec{r}'\rangle}c^\dagger_{\vec{r}}c_{\vec{r}'}-\mu\sum_{\vec{r}}c^\dagger_{\vec{r}}c_{\vec{r}},
\end{eqnarray}
where $t$ is a real positive number with chemical potential $\mu$ and $c^\dagger_{\vec{r}}$ the creation operator of spinless fermion at lattice site $\vec{r}$ with $\langle\cdots\rangle$ denoting summations only over nearest-neighboring sites.

By a generalized 't Hooft anomaly matching~\cite{Cho:2017aa,Metlitski:2018aa,Yao:2019aa,Hongo:2019aa}, nontrivial symmetry anomalies in a low-energy effective field theory of an arbitrarily fine-tuned lattice model implies an ingappability of general lattice models, constrained only by filling and symmetry structures. Therefore, the following analysis for the non-interacting model should be also valid for more general, interacting systems.

Under the TLBC,
we can exhaust all the sites by translating a single site to one direction only, e.g. $\vec{r}=j\hat{x}_1$ with $j=1,2,\cdots,V$. With this coordinate convention, we further define $\Psi_j\equiv c_{j\hat{x}_1}$ which implies
\begin{eqnarray}
\label{T-B}
H_\text{T-B}&=&-t\sum_{j=1}^V\left[\sum_{i=1}^{d}\Psi^\dagger_{j+V_i}\Psi_{j}+\text{h.c.}\right]-\mu\sum_{j=1}^V\Psi^\dagger_j\Psi_j;\nonumber\\
V_i&\equiv&\frac{V}{\prod_{j=i}^{d}L_j},
\end{eqnarray}
with the boundary condition as $\Psi_j=\Psi_{j+V}$ if the range of $j$ is extended to all integers. Then we can express the Hamiltonian in the momentum space by
\begin{eqnarray}
\label{fourierexp}
\Psi_j=\frac{1}{\sqrt{V}}\sum_{k=0}^{V-1}\Psi(k)\exp\left(- i \frac{2\pi k}{V}j\right),
\end{eqnarray}
as
\begin{eqnarray}
H_\text{T-B}&=&\sum_{k=0}^{V-1}\left\{-2t\left[\sum_{i=1}^d\cos\left(\frac{2\pi V_i}{V}k\right)\right]-\mu\right\}\Psi^\dagger(k)\Psi(k)\nonumber\\
&=&\sum_{k\in\text{B.Z.}}\left\{-2t\left[\sum_{i=1}^d\cos\left(\frac{2\pi V_i}{V}k\right)\right]-\mu\right\}\Psi^\dagger(k)\Psi(k)\nonumber\\
&\equiv&\sum_{k\in\text{B.Z.}}\epsilon(k)\Psi^\dagger(k)\Psi(k),
\end{eqnarray}
where we have chosen the Brillouin zone which is symmetric at $k=0$ so that the zeros of $\epsilon(k)$ is symmetric around the origin. Thus we can label these zeros as $\{\pm K_c\}_{c=1,\cdots,N_0}$ due to the even parity: $\epsilon(k)=\epsilon(-k)$.
{The low-energy Hamiltonian, which describes excitations near
the zeros of $\epsilon(k)$, is given as}
\begin{eqnarray}
\mathscr{H}=\sum_{c=1}^{N_0}\int_{-\Lambda<k<\Lambda}v_ck\left(\Psi^\dagger_c(k)\Psi_c(k)-\Psi^\dagger_{\bar{c}}(k)\Psi_{\bar{c}}(k)\right). \nonumber\\
\label{eq.Heff}
\end{eqnarray}
Thus $N_0 \sim L_2 \times L_3 \ldots \times L_d$
represents the number of the one-dimensional channels ($1+1$-dimensional Dirac fermions).
{Here} the ultraviolet cut-off $\Lambda\ll2\pi/(N_0a)$, where $a$ is the lattice constant,
and $v_f$ is the value of fermi velocity not necessarily positive. The fermionic operator $\Psi_c(k)=\Psi(K_c+k)$ and $\Psi_{\bar{c}}(k)=\Psi(-K_c+k)$.
It is essential to remark the role played by the charge quantization condition in Eq.~(\ref{charge_quantization}).
{To obtain a free system in the continuum limit where $k$ is taken to be a continuum variable on the whole real axis, it is necessary that the $k=0$ modes or $\Psi(\pm K_c)$ modes exist in the Hilbert space under the continuum limit, as we will see later.}
In $d=1$, Eq.~(\ref{charge_quantization}) implies precisely such existence of $\Psi(\pm K_c)$ modes. Therefore, we expect this conclusion generalizes to arbitrary $d\geq1$ and the charge quantization indeed permits us to do this continuum limit.

We absorb the $|v_c|$ into the following definition of two-component Dirac operator:
\begin{eqnarray}
\label{dirac}
\psi_c(k)=\left\{\begin{array}{ll}{\left(\begin{array}{c}|v_c|\Psi_c(k)\\|v_c|\Psi_{\bar{c}}(k)\end{array}\right)},&v_c>0;\\{\left(\begin{array}{c}|v_c|\Psi_{\bar{c}}(k)\\|v_c|\Psi_c(k)\end{array}\right)},&v_c<0. \end{array}\right.
\end{eqnarray}
With the definition above, the real space formulation of Lagrangian density takes a compact form as
\begin{eqnarray}
\label{boundary_dirac}
\mathscr{L}_{U(1)}=\sum_{c=1}^{N_0}\sum_{\mu=0}^1\bar{\psi}_c(t,x)i\gamma^\mu\partial_\mu\psi_c(t,x),
\end{eqnarray}
where $\gamma^0=\sigma_2$ and $\gamma^1=\sigma_1$ with $\vec{\sigma}$ Pauli matrices, and the chirality is $\gamma_3=\sigma_3$.
{
The role played by the existence of the zero mode $\psi_{c}(k=0)$ or $\Psi(\pm K_c)$ can be understood in the real space formulation (\ref{boundary_dirac}): if we did not have these zero modes, we would discard the zero mode of the Fourier expansion of $\psi_c(t,x)$ by hand when we quantize it, which makes (\ref{boundary_dirac}) not free Dirac fermions.
}

Then we minimally couple it with an external gauge field as
\begin{equation}
A_1(t,x)= \frac{2\pi t}{VT} ,
\label{eq.defA1}
\end{equation}
and $A_0(t,x)=0$: $\partial_\mu\rightarrow\partial_\mu-ieA_\mu$,
which exactly corresponds to the flux insertion defined by Eq.~(\ref{flux}) except for that we are left by only one spatial dimension since we have applied the coordinate convention $\vec{r}=j\hat{x}_1$ with $j=1,\cdots,V$ before on the lattice.

Combining Eqs.~(\ref{fourierexp},\ref{dirac}), we obtain the lattice translation $T_1$ in $\hat{x}_1$ direction representation of $\psi_c(t,x)$ as:
\begin{eqnarray}
\label{chiral}
\psi^{T_1}_c(t,x)&\equiv& T_1\psi_c(t,x)T_1^{-1}\nonumber\\
&=&\exp\left[ i \text{sgn}(v_c)\pi\gamma_3\frac{2K_c}{V}\right]\psi_c(t,x),
\end{eqnarray}
which is simply the chiral symmetry transformation, at low energy, appearing on-site. It is straightforward to apply Fujikawa's method~\cite{Fujikawa:2004aa,Fujikawa:2004ab} to calculate the global symmetry anomaly as the phase ambiguity of the fermionic partition function responding to the chiral transformation as Eq.~(\ref{chiral}) in the background gauge field configuration ${A}_\mu(t,x)$.
\begin{eqnarray}
\label{partition_chiral}
&&Z_{U(1)}\nonumber\\
&\equiv&\frac{\int\mathscr{D}(\bar{\psi}^{T_1},\psi^{T_1})\exp(-\int \mathscr{L}_{U(1)}[\bar{\psi}^{T_1}_c(\tau,x),\psi^{T_1}_c(\tau,x),{A}_\mu])}{\int\mathscr{D}(\bar{\psi},\psi)\exp(-\int \mathscr{L}_{U(1)}[\bar{\psi}_c(\tau,x),\psi_c(\tau,x),{A}_\mu])}\nonumber\\
&=&\exp\left[ i \sum_{c=1}^{N_0}2\text{sgn}(v_c)\frac{2\pi K_c}{V}\right]\nonumber\\
&=&\exp( i 2\pi\nu),
\end{eqnarray}
where the fermionic measure $\mathscr{D}(\bar{\psi}^{T_1},\psi^{T_1})=J_{T_1}\cdot\mathscr{D}(\bar{\psi},\psi)$ is calculated by Fujikawa's $U(1)$ gauge-invariant regularization~\cite{Fujikawa:2004aa,Fujikawa:2004ab} and we have used the fact the filling fraction is $\nu$ and $\nu=\sum_c2\text{sgn}(v_c)K_c/V$, and evaluate the formal path integral in the Euclidean signature: $\tau=it$.

We define a lattice partition function ratio as:
\begin{eqnarray}
\label{partition_lattice}
Z^{\mbox{\scriptsize latt}}_{U(1)}&\equiv&\frac{\text{Tr}_\text{G.S.}\left[T_1\hat{U}_0\hat{U}_\text{flux}T_1^{-1}\right]}{\text{Tr}_\text{G.S.}\left[\hat{U}_0\hat{U}_\text{flux}\right]}\nonumber\\
&=&\frac{\langle\text{G.S.}|T_1\hat{U}_0\hat{U}_\text{flux}T_1^{-1}|\text{G.S.}\rangle}{\langle\text{G.S.}|\hat{U}_0\hat{U}_\text{flux}|\text{G.S.}\rangle},
\end{eqnarray}
where we denote by ``$\text{Tr}_\text{G.S.}$'' taking the trace only within lowest energy states since the Wick rotation ``$\tau=it$'' in the definition of $Z_{U(1)}$ implies that we should project out excited states when we define $Z^{\mbox{\scriptsize latt}}_{U(1)}$. We also assume a unique gapped lattice ground state $|\text{G.S.}\rangle$ and $\hat{U}_\text{flux}$ denotes the unitary time evolution by flux insertion followed by the large gauge transformation $\hat{U}_0$.
The continuum-limit form of $Z'_\text{U(1)}$ in Eq.~(\ref{partition_lattice}) exactly coincides with the form of $Z_{U(1)}$ in Eq.~(\ref{partition_chiral}), where the large gauge transformation $\hat{U}_0$ is implicitly presented in the path integrals within $Z_{U(1)}$ since the inner product of wave functionals $|\{\bar{\psi},\psi\}\rangle_{{A}_\mu}$'s at the last time slice can be done only after fixing the gauge by the (large) gauge transformation $\hat{U}_0$, which is in the same situation as the lattice model.
The necessity of $\hat{U}_0$ can be seen once one notices that the wave functional $\{\psi\}$ is an associated complex line sector of the underlying $U(1)_Q$ principal bundle and $\hat{U}_0$ in Eq.~(\ref{large_U}) is exactly the gluing transition function between the initial and the final time slices.
It implies that $Z_{U(1)}$ is the low-energy limit of $Z^{\mbox{\scriptsize latt}}_{U(1)}$.

{
Then we can rephrase the consequence of anomaly $Z_{U(1)}$ that the (discrete) chiral symmetry (\ref{chiral}) will be broken once $U(1)_Q$ is gauged, in the lattice language.
Since the anomalous chiral symmetry means the chiral charge is not conserved and here the chiral symmetry corresponds to the lattice translation, we expect the lattice momentum is not conserved
under the gauge field.
Indeed, this is a well known phenomenon that the particles are
accelerated by electric field and acquire a momentum.
However, it should be noted that the momentum is a gauge-dependent
quantity.
Under the time-dependent uniform gauge field~\eqref{eq.defA1}, the
momentum is exactly conserved thanks to the translation invariance.
However, after a finite time, the gauge field (vector potential) is nonzero
and the momentum cannot be directly compared to the initial value.
Nevertheless, when the system encloses an integral multiple of the
unit flux quantum ($2\pi$), the gauge field can be exactly
eliminated by the large gauge transformation $\hat{U}_0$, so that
the momentum can be compared with its initial value.
It is this large gauge transformation $\hat{U}_0$ that induces
the change in the momentum, as we can see from the flux-insertion
argument for the LSM theorem~\cite{Oshikawa:2000aa}.
This observation also supports the identification of $Z_{U(1)}$
as the low-energy limit of $Z^{\mbox{\scriptsize latt}}_{U(1)}$.
{Additionally, the role of the gauge invariance of Fujikawa's regularization applied in the calculation of Eq.~(\ref{partition_chiral}) should be noted since a gauge non-invariant regularization method can give a vanishing chiral anomaly~\cite{Peskin:2018aa,Yao:2019ab} corresponding to the conservation of the lattice momentum in the fixed initial gauge, which can be also explicitly seen by the translational symmetry of the lattice Hamiltonian.
It exactly reflects the mixing nature of the chiral anomaly, which characterizes the conflicting between $U(1)$ and $T_1$. }
}

Since $Z_{U(1)}$ is a topological invariant, e.g. invariant along any symmetry-respecting renormalization-group flow by the generalized 't Hooft anomaly-matching, we can evaluate $Z^{\mbox{\scriptsize latt}}_{U(1)}$ by its low-energy limit $Z_{U(1)}$:
\begin{eqnarray}
Z^{\mbox{\scriptsize latt}}_{U(1)}=Z_{U(1)}.
\end{eqnarray}
However, the unique ground state must be featureless hence a $T_1$-eigenstate, which implies $T_1|\text{G.S.}\rangle=\exp(iP)|\text{G.S.}\rangle$ thereby $Z_{U(1)}=1$. This contradicts with $Z_{U(1)}=Z^{\mbox{\scriptsize latt}}_{U(1)}=\exp(i2\pi\nu)$ if $q\neq1$, unless
\begin{eqnarray}
\hat{U}_\text{flux}|\text{G.S.}\rangle\perp|\text{G.S.}\rangle,
\end{eqnarray}
which still conflicts the unique gapped ground state. Thus the ground states must be degenerate for fractional fillings.
Then we arrive at the LSM theorem with a well-defined anomaly-manifestation in a general thermodynamic limit.
Furthermore, the physical interpretation of $Z_{U(1)}$ can be understood by its lattice partner $Z^{\mbox{\scriptsize latt}}_{U(1)}$ which exactly measures the momentum changes after the charge pumping. In this sense, we call that the chiral anomaly derived from the Dirac field theory is lattice-realized.
The anomaly we have considered is not an emergent anomaly at low energy~\cite{Cho:2017aa,Metlitski:2018aa} which can be seen as follows.

Let us first tune the hoppings in (\ref{T-B}) nonzero only along $\hat{x}_1$.
At low energy under TLBC, $U(1)$ and translation symmetries (\ref{chiral}) are reduced to $U(1)\times\mathbb{Z}_q$~\cite{Cho:2017aa}, where $\nu=p/q$ defined before.
Since the translation is actually represented by $\mathbb{Z}$ on the lattice with TLBC in the thermodynamic limit, we should do a symmetry extension,
which is a quantitative treatment of the translation symmetry at low-energy honestly as $\mathbb{Z}$ rather than its emergent expression $\mathbb{Z}_q$~\cite{Cho:2017aa}. The symmetry extension is a mapping from the $[U(1)\times\mathbb{Z}_q]$-anomaly classes to the $[U(1)\times\mathbb{Z}]$-anomaly classes, associated with the dual of the third mapping in the short exact sequence $0\rightarrow\mathbb{Z}\xrightarrow{q}\mathbb{Z}\rightarrow\mathbb{Z}_q\rightarrow0$ of symmetries~\footnote{Here the associated dual mapping is $\Omega^{3}_{\text{Spin}^c}(B\mathbb{Z}_q)\rightarrow\Omega^{3}_{\text{Spin}^c}(B\mathbb{Z})$, where $\Omega$ is the spin$^c$ cobordism of the classifying spaces $B\mathbb{Z}_q$ and $B\mathbb{Z}$. $U(1)$ gauge contribution has been taken into account by the spin$^c$ structure.},
This symmetry extension does not trivialize
the anomaly classes generating by (\ref{partition_chiral})~\footnote{The mixed anomaly classes generated by the anomaly factor $\exp(i2\pi\nu)$ forms $\mathbb{Z}_q\subset\Omega^3_{\text{Spin}^c}(B\mathbb{Z}_q)$ and the symmetry extension maps this $\mathbb{Z}_q$ injectively to $\Omega^3_{\text{Spin}^c}(B\mathbb{Z})\cong U(1)$ by the natural inclusion $\mathbb{Z}_q\xhookrightarrow{}U(1)$. }, which means the anomaly classes generated by (\ref{partition_chiral}) are still there even when we treat translation faithfully.
The general cases where the translation is represented by (\ref{chiral}) can be argued by anomaly-matching to the fine-tuned point above.
In contrast, if we orbifolded theory by the $\mathbb{Z}_q$ symmetry using its finite cyclicity, which no longer holds at the lattice scale, more anomalies would emerge, but eventually be trivialized after the symmetry extension~\cite{Cho:2017aa}.
{Furthermore, we calculate the anomaly at a critical point with a special translation representation (\ref{chiral}), which can be expected as an intrinsic anomaly at the lattice level even if the lattice model is tuned away from this point, in the following way. Let us assume the lattice degrees of freedom are realized by a higher dimensional lattice bulk $B$ with the corresponding $U(1)\times\mathbb{Z}$ symmetry at the lattice scale.
Then the intrinsic anomaly is unambiguously determined by the $(U(1)\times\mathbb{Z})$-crystalline-SPT (cSPT) class that this bulk $B$ belongs to.
It has been shown that cSPT one-to-one corresponds to an SPT where $U(1)\times\mathbb{Z}$ is realized purely on-site and the continuum limit here is proposed to give such a correspondence~\cite{Metlitski:2018aa,Huang:2017aa,Thorngren:2018aa}.
It means this (onsite) SPT class corresponds to the anomaly of our critical point in the continuum limit, calculated by the representation (\ref{chiral}) in a $\mathbb{Z}$ sense.
Then the anomaly (\ref{partition_chiral}) is associated to that cSPT, thereby intrinsic and applicable away from the chosen critical point.}

Finally, for $d=0$, due to Eq.~(\ref{charge_quantization}), the filling $\nu_0\in\mathbb{N}$. Then, the low-energy effective response theory is simply the $(0+1)$-dimensional Chern-Simons theory with level $\nu_0$:
\begin{eqnarray}
s_{U(1)}=\nu_0\int dtA(t),
\end{eqnarray}
by a minimal coupling observation. Such an effective action is well-defined since $\nu_0$ is an integer, and the zero-dimensional theory is $U(1)_Q$ anomaly-free thereby well-defined. It is applicable even for bosonic theory since there is no spatial coordinate to give one a choice of spin structures while the boundary condition along $t$ or $\tau$ for field operator is already fixed as, respectively, periodic and anti-periodic for bosonic and fermionic situations. Thus, $d=0$ cases can be always trivially gapped if we only have a $U(1)_Q$ symmetry.
This agrees with the trivial $d=0$ version of the LSM theorem.
While the statement is rather trivial, this is still a useful exercise to
check the consistency of the anomaly argument.
In the next section, we will see somewhat more non-trivial consistency check in $d=0$ for a higher symmetry.

\subsection{Bulk-boundary correspondences: LSM theorem with $U(1)_Q$}
\label{bulk-boundary_u(1)}
{In this Subsection, based on our field-theory formulation related to the TLBC, we discuss the LSM theorem from the point of view
of anomaly inflow.
This enables us to construct
a higher dimensional SPT bulk theory where the massless theory $\mathscr{L}_{U(1)}$ in Eq.~(\ref{boundary_dirac}) can be seen as a boundary theory attached to
the SPT bulk.}
{
To simplify the discussion, in the following we first assume $N_0=1$.
We will generalize the argument to arbitrary $N_0$ later.}
The translation symmetry is reduced to:
\begin{eqnarray}
\label{transl_edge}
\psi^{T_1}&=&\exp\left(i\gamma_3\pi\nu\right)\psi=\exp\left(i\gamma_3\pi\frac{p}{q}\right)\psi.
\end{eqnarray}
{
Our physical degrees of freedom is the free Dirac fermion composed by $\psi=[\psi_\text{L},\psi_\text{R}]$ where $\psi_\text{L,R}$ are the left- and right-moving chiral fermions, respectively.
It is well-known that a single left-moving chiral fermion can be generated on the edge of an integer quantum Hall system with $\sigma_\text{H}=+1$ and a right mover by $\sigma_\text{H}=-1$.
Thus the (bulk) matter field, which is able to support $\psi_\text{L}$ and $\psi_\text{R}$ fermions on its edge, consists of two $(2+1)$-dimensional massive Dirac fermion with opposite masses, which realizes $\sigma_\text{H}=\pm1$, separately~\footnote{The Pauli-Villars regulator of each massive Dirac fermion has an opposite mass to the regulated fermion.}.
Then we couple the background $U(1)_Q$ gauge field $A_{U(1)}$ and the gauge field $a\gamma^3/2$ of the translational symmetry in (\ref{transl_edge}) to the bulk fermions.}
After the bulk matter field is integrated out,
\begin{eqnarray}
S_{\text{bulk},U(1)}=\int
\left[
-\frac{i}{4\pi}A_\text{L} \wedge dA_\text{L}+
\frac{i}{4\pi}A_\text{R}\wedge dA_\text{R}\right], \nonumber\\
\end{eqnarray}
where
\begin{eqnarray}
\label{chiral_gauge}
A_\text{L}&=&A_{U(1)}+a/2;\,\,\,A_\text{R}=A_{U(1)}-a/2.
\end{eqnarray}
where we have taken a ``$1/2$'' normalization convention so that $a$ has the same constraint as a $\mathbb{Z}_q$-gauge field:
\begin{eqnarray}
\label{z_q_constraint}
da=0\,\,\text{ and }\oint_\text{closed loop}a\in2\pi\frac{p}{q}\mathbb{Z}.
\end{eqnarray}
Then,
{
\begin{eqnarray}
\label{bf_u(1)}
S_{\text{bulk},U(1)}=-\int\frac{i}{2\pi}a \wedge dA_\text{U(1)},
\end{eqnarray}
}
which is exactly a BF-theory~\cite{Horowitz:1989aa,Maldacena:2001aa} coupling the $\mathbb{Z}_q$-gauge field and the $U(1)_Q$-gauge field. To detect the nontrivial aspect of such an SPT bulk, we evaluate the bulk partition function on a compact closed manifold, e.g. three-torus $T^3=T^2\times S^1$, with the following background gauge field:
\begin{eqnarray}
\label{background_u(1)}
&&\int_{T^2}\frac{dA_{U(1)}}{2\pi}=1,\,\,\int_{S^1}a=2\pi\nu,
\end{eqnarray}
where $a$ is a pull-back from a flat gauge field on $S^1$ and $A_{U(1)}$ is a pull-back from a gauge field on $T^2$ with a unit Chern number.
With this background gauge field configuration, we obtain the bulk partition function as
\begin{eqnarray}
Z_{\text{bulk},U(1)}&=&\exp(-S_{\text{bulk},U(1)})\nonumber\\
&=&\exp(i2\pi\nu).
\end{eqnarray}
{To understand the physical meaning of this bulk partition function evaluated on $T^2\times S^1$, we briefly review the useful tool called mapping torus~\cite{Atiyah:1975ab,Callan-Jr:1978aa,Kiskis:1978aa,Witten:1985aa,Witten:2016aa} below.
Given a potentially anomalous theory defined on a physical spacetime $\mathcal{M}$, its phase ambiguity of partition function by a transformation, e.g. a gauge or global symmetry transformation, can be calculated by the partition function of corresponding bulk on a manifold $\mathcal{M}\times S^1$~\cite{Witten:1985aa,Witten:2016aa}  constructed in the following way.
We take a ``cylinder'' $\mathcal{M}\times [0,1]$ where the bulk field stays and it reproduces the pre-transformed boundary theory at $\mathcal{M}\times\{0\}$ and the transformed theory at the other end $\mathcal{M}\times\{1\}$.
Then we paste these two ends by the certain transformation twisting, which corresponds to a twisted boundary condition of the bulk field on $\mathcal{M}\times I$, to form the mapping ``torus'' $\mathcal{M}\times S^1$.
In our case (\ref{background_u(1)}) above, we take $S^1$ as the extra dimension and $T^2$ represents the physical space-time of boundary theory.
The holonomy $\int_{S^1}a$ precisely realizes the twisted $T_1$ transformation of the bulk field and the instanton $\int_{T^2}dA_{U(1)}/2\pi$ implies the flux insertion process by a unit flux.
Therefore, this mapping-torus bulk partition function is equals to the phase ambiguity, brought by a $T_1$ transformation, of the boundary partition function with a flux insertion in physical spacetime, which is precisely the quantity calculated in (\ref{partition_chiral}). }

This bulk construction can be generalized to arbitrary $N_0$ in $\mathscr{L}_{U(1)}$ by replacements of ``$1/2$'' in Eq.~(\ref{chiral_gauge}) by color-dependent coefficients ``$\text{sgn}(v_c)K_c/(V\nu)$'' due to Eq.~(\ref{chiral}). Then, after the color indices ``$c$'' is summed up and by $\nu=\sum_c2\text{sgn}(v_c)K_c/V$, we arrive at the same effective bulk response theory as Eq.~(\ref{bf_u(1)}).

{
The bulk description of the anomaly proposes an alternative way to detect the anomaly by the boundary partition function~(See e.g. Appendix~D of \cite{Song:2019aa}).
Let us take the $T^2$ in (\ref{background_u(1)}) as the manifold spanned by the physical spatial dimension and the extra dimension. $S^1$ is taken as the physical dimension of time.
Then the holonomy $\left(\int_{S^1}a\right)$ means we translate the system by a lattice distance during the time period $S^1$ and thus the resultant partition function is an exponential of the lattice momentum.
The instanton in $T^2$ means that we are comparing this momentum obtained in two gauge choices of the gauge field along the physical spatial dimension,
which are connected by a large gauge transformation.

Therefore, the bulk-response approach (\ref{bf_u(1)}) provides a unified way to describe its boundary anomaly.
}

\section{LSM theorem for $SU(N)$ spin-rotation symmetry}
\label{LSMA}

In the following, we will use the methods developed in the previous section
 to investigate the higher symmetry generalization of LSM theorem, e.g. replacement of the global onsite symmetry by $SU(N)$ spin-rotation symmetry, or more precisely a $PSU(N)=SU(N)/\mathbb{Z}_N$ global symmetry~\cite{Cheng:2016aa,Yao:2019aa} by the spin operator satisfying the following $su(N)$ Lie algebra commutation relations:
\begin{eqnarray}
\label{su(n)_algebra}
\left[S^\alpha_{\vec{r},\beta},S^\gamma_{\vec{r}',\delta}\right]=\delta_{\vec{r},\vec{r}'}\left(\delta^\alpha_\delta S^\gamma_{\vec{r},\beta}-\delta^\gamma_\beta S^\alpha_{\vec{r},\delta}\right),
\end{eqnarray}
where $\alpha$ and $\beta$ are the ``spin'' indices that take values among $1$ to $N$.
In particular, we focus on the quantum-anomaly manifestation of the LSM theorem.
Its advantage is that the lattice ingappabilities can be detected at any fine-tuned critical point which simplifies the calculation, thanks to the 't Hooft anomaly matching.

\subsection{``Spin''-quantization condition}
Let us consider the most general situation that the total number of Young-tableaux boxes per unit cell is $b$. Analogous to Eq.~(\ref{charge_quantization}), we also take the following ``spin''-quantization condition:
\begin{eqnarray}
\label{spin_quantization}
{b}V/N\in\mathbb{N}.
\end{eqnarray}
The reason we assume this condition is different from the $U(1)_Q$ LSM case where charge quantization is naturally imposed by fundamental degrees of freedom. To clarify this seemingly unnatural requirement, let us see the situation where Eq.~(\ref{spin_quantization}) is not satisfied, namely ${b}V/N\notin\mathbb{N}$. Then the total Young-tableaux boxes $bV$ of the system is not divisible by $N$. By the knowledge from representation theory, there is no $SU(N)$-singlet sector in the Hilbert space and by the global $PSU(N)$ symmetry, the system is exactly degenerate, even at all excited states since the states within any non-singlet irreducible representation must have the same energy due to Schur's lemma applied within any of these nontrivial irreducible sectors.

Such a rather trivial type of ingappabilities is essential to understand finite-system spectrum. Nevertheless, we are not interested in it since these ingappabilities are not a many-body effect and they depend on a specific choice of system sizes. A many-body ingappability is commonly considered as almost degeneracies, e.g. the ingappabilities make sense only at general thermodynamic limits.
Therefore, in the following discussion for $(d\geq1)$-dimensional system, we avoid such an exact degeneracy explicitly exposed above by imposing Eq.~(\ref{spin_quantization}). Nevertheless, when stating the final theorem without loss of generality, we will also include the cases that $bV/N\notin\mathbb{N}$ for which the ground states are exactly degenerate.

\subsection{Generalized LSM theorem and anomaly manifestation}
Since we are only interested in the low-energy spectrum of the lattice model, we can equivalently reconstruct its low-energy physical properties by coupling $b$ copies of the lattice models each of which has one fundamental $SU(N)$ degree of freedom within each unit cell. It is possible due to the group-theoretical knowledge that any $SU(N)$ irreducible representation with $b_0$ Young-tableaux boxes is contained in the tensor products of $b_0$ of fundamental representations. Then we can project out all the undesired degrees of freedom in the analog of Affleck-Kennedy-Lieb-Tasaki chain construction from spin-$1/2$ degrees of freedom~\cite{Affleck:2004aa}. Moreover, such a projection can be realized by a strong interaction dynamically, hence the quantum anomaly factor of the original system can be obtained by a summation of the anomalies of these $b$ of fundamental lattices. Then the problem is reduced to the anomaly related to fundamental lattices we will calculate below.

Similarly to the generalized LSM theorem in one dimension, we do the following $N$-flavor fermionization representing spin degrees of freedom:
\begin{eqnarray}
\label{spinor}
S^\alpha_{\vec{r},\beta}=\Psi^{\alpha\dagger}(\vec{r})\Psi_{\beta}(\vec{r})-\frac{1}{N}\delta^\alpha_\beta
\end{eqnarray}
with the restriction of total particle number on every site ``$\vec{r}$''
\begin{eqnarray}
\label{restriction}
\sum_{\alpha=1}^N\Psi^{\alpha\dagger}(\vec{r})\Psi_{\alpha}(\vec{r})=1,
\end{eqnarray}
so that $S^\alpha_{\vec{r},\beta}$'s defined in Eq.~(\ref{spinor}) satisfy Eq.~(\ref{su(n)_algebra}).

Again, we impose the TLBC defined as Eq.~(\ref{tbc0}). In the analog of $SU(2)$ cases, a fine-tuned critical model can be the Hubbard model in the conductive phase:
\begin{eqnarray}
\label{Hubbard}
H_{SU(N)}&=&-t\sum_{\langle\vec{r},\vec{r}'\rangle}\left[\sum_{\alpha=1}^{N}\Psi^{\alpha\dagger}(\vec{r})\Psi_\alpha(\vec{r}')+\text{h.c.}\right]\nonumber\\
&&-U\left[\sum_{\alpha=1}^N\Psi^{\alpha\dagger}(\vec{r})\Psi_\alpha(\vec{r}')-1\right]^2,
\end{eqnarray}
where the Hund's rule coupling $U\gg|t|$ realizes the particle number restriction per unit cell in Eq.~(\ref{restriction}). By an observation on Eq.~(\ref{Hubbard}) and Eq.~(\ref{T-B}), we can see that: 1) $SU(N)$ case has a stronger particle number restriction per unit cell; 2) each flavor has a filling fraction $1/N$ within a unit cell. With this comparison, it is straightforward to derive the effective theory of the current $SU(N)$ model after we apply the coordinate system $\vec{r}=j\hat{x}_1$ with $j=1,\cdots,V$ again:
\begin{eqnarray}
\mathscr{L}_{SU(N)}[{\mathcal A}]&=&\sum_{c=1}^{N_0}\sum_{\alpha=1}^N\bar{\psi}^\alpha_c(t,x)i\gamma^\mu(\partial_\mu-i{\mathcal A}_\mu)\psi_{c,\alpha}(t,x), \nonumber\\
\label{effective_filling}
\nu_\text{eff}&=&\frac{1}{N},
\end{eqnarray}
where the inclusion of a dynamical fluctuating $U(1)_Q$ gauge field is used to eliminate $U(1)_Q$ phase degrees of freedom. It is because $U(1)_Q$ is unphysical, e.g. the fundamental observable $S^\alpha_{\vec{r},\beta}$ is invariant under $U(1)_Q$.
It can be also viewed as a realization of the particle number constraint as Eq.~(\ref{restriction}) regularized by the Grassmann-number ordering ambiguity to make the Lagrangian explicitly $U(1)_Q$ gauge invariant:
\begin{eqnarray}
\begin{array}{c}\sum_{\alpha=1}^N\Psi^{\alpha\dagger}(\vec{r})\Psi_{\alpha}(\vec{r})=1,\\\downarrow\\\sum_{\alpha=1}^N\left(1-\frac{1}{N}\right)\Psi^{\alpha\dagger}(\vec{r})\Psi_\alpha(\vec{r})-\frac{1}{N}\Psi_\alpha(\vec{r})\Psi^{\alpha\dagger}(\vec{r})=0,\\\downarrow_\text{Grassmanian}\\\sum_\alpha\left(1-\frac{1}{N}\right)\psi^{\alpha\dagger}(t,x)\psi_\alpha(t,x)-\frac{1}{N}\psi_\alpha(t,x)\psi^{\alpha\dagger}(t,x)=0,\\\downarrow_\text{Reordering}\\\sum_\alpha\psi^{\alpha\dagger}(t,x)\psi_\alpha(t,x)=0, \end{array}\nonumber
\end{eqnarray}
which can be done by a Lagrangian multiplier $\sum_{\alpha}{\mathcal A}_0\bar{\psi}^\alpha\gamma^0\psi_\alpha$ consistent with gauge invariance.
The zeros of Hamiltonian $\epsilon_\alpha(k)$ for each flavor in momentum space in the same notations satisfy:
\begin{eqnarray}
\sum_{c=1}^{N_0}2\text{sgn}(v_c)K_c/V=\nu_\text{eff}.
\end{eqnarray}
Similarly to Eq.~(\ref{charge_quantization}), the $U(1)_Q$-LSM case, Eq.~(\ref{spin_quantization}) also guarantees the existence of the continuum limit of $k$ variables in $\Psi_{\alpha}(\pm K_c+k)$.
Then we couple the theory to a background $PSU(N)$ gauge field $A_{PSU(N)}$ which locally takes value in $su(N)$ algebra.
In the following discussion, $t_{N^2-1}\equiv\text{diag}[1,1,\cdots,1,-(N-1)]_{N\times N}$ is denoted as the matrix representation of the last $su(N)$ generator.
First, let us analyze the current gauge group, which actually is not $U(1)\times SU(N)=\{(g_1,g_N)|g_1\in U(1),g_N\in SU(N)\}$ since its center $\mathbb{Z}_N$ generated by $(\exp(i2\pi/N),\exp(-i2\pi t_{N^2-1}/N))\in U(1)\times SU(N)$ does not transform the matter field $\{\psi_\alpha\}$.
It implies that the gauge group is $[U(1)\times SU(N)]/\mathbb{Z}_N\cong U(N)$, and
{
thus only $A_{PSU(N)}+{\mathcal A}$, $NA_{PSU(N)}$ and $N{\mathcal A}$ or their integer multiples can be lifted to canonical 1-form $u(N)$ connections thereby expressible by well-defined $u(N)$ connections.}

We consider the following flux insertion which generalizes the $U(1)$ case and {the connection can be written down by a $u(N)$ connection}:
\begin{eqnarray}
\label{extension_psu(N)}
A_{PSU(N)}+{\mathcal A}=\frac{2\pi t}{TVN}(1-t_{N^2-1})_{N\times N}+\delta{\mathcal A},
\end{eqnarray}
where $\delta{\mathcal A}$ is a dynamical canonical $U(1)$ connection since $\left[{2\pi t}(1-t_{N^2-1})/(TVN)\right]$ is globally well-defined, and then the functional integration $\int\mathscr{D}{\mathcal A}$ over $u(1)/\mathbb{Z}_N$ connection can be converted to $\int\mathscr{D}\delta{\mathcal A}$ over $u(1)$ connection.
The physical interpretation of such a $PSU(N)$ ``flux'' insertion is that we adiabatically rotate the spin along $\hat{x}_1$ direction and this spatial dependent rotation matrix projectively represented by $SU(N)$ matrix is multi-valued since it is identified up to a center of $SU(N)$ due to the local $U(1)$ gauge degrees of freedom by ${\mathcal A}$. Therefore, the accompany $U(1)$ twisting is simply to compensate this artificial ambiguity in the $SU(N)$ rotation matrix.

In a similar sense, we can calculate the anomaly factor by Fujikawa's gauge-invariant regularization~\cite{Fujikawa:2004aa,Yao:2019aa} on the fermionic measure $\prod_{c,\alpha}\mathscr{D}(\bar{\psi}_{c,\alpha}^{T_1},\psi_{c,\alpha}^{T_1})=J(T_1)\prod_{c,\alpha}\mathscr{D}(\bar{\psi}_{c,\alpha},\psi_{c,\alpha})$ below:
\begin{eqnarray}
\label{partition_su(N)}
&&Z_{SU(N)}\nonumber\\
&\equiv&\frac{\int\mathscr{D}{\mathcal A}\mathscr{D}(\bar{\psi}_{c,\alpha}^{T_1},\psi_{c,\alpha}^{T_1})\exp(-\int \mathscr{L}^{T_1}_{SU(N)}[{\mathcal A}+{A}_{PSU(N)}])}{\int\mathscr{D}{\mathcal A}\mathscr{D}(\bar{\psi}_{c,\alpha},\psi_{c,\alpha})\exp(-\int \mathscr{L}_{SU(N)}[{\mathcal A}+{A}_{PSU(N)}])}\nonumber\\
&=&\exp\left[ i \sum_{c=1}^{N_0}2\text{sgn}(v_c)\frac{2\pi K_c}{V}\right]\nonumber\\
&=&\exp\left( i{2\pi}/{N}\right),
\end{eqnarray}
where we have made use of the same notation for the transformation rule of $\psi\rightarrow\psi^{T_1}$ as Eq.~(\ref{chiral}) and $\mathscr{L}^{T_1}_{SU(N)}\equiv\mathscr{L}_{SU(N)}[\bar{\psi}^{T_1}(\tau,x),\psi^{T_1}(\tau,x)]$ except that here we have a specified filling factor $\nu_\text{eff}=1/N$ for each flavor.

Thus for the general case of $b$ of Young-tableaux boxes per unit cell:
\begin{eqnarray}
\label{ano_su(N)}
Z^{(b)}_{SU(N)}&=&\left(Z_{SU(N)}\right)^b=\exp\left( i 2\pi\frac{b}{N}\right).
\end{eqnarray}
Similarly to the $U(1)$ case in LSM theorem, we can also define the lattice-realization of $Z^{(b)}_{SU(N)}$ by ${Z'}^{(b)}_{SU(N)}$ analogous to Eq.~(\ref{partition_lattice}), and by anomaly-matching: ${Z'}^{(b)}_{SU(N)}=Z^{(b)}_{SU(N)}$.

\subsection{LSM-type anomaly in $0$ dimension}

{
To complete the $SU(N)$ generalization of LSM theorem, we also study the case of $d=0$, which is a problem in single-body quantum mechanics.
Unlike in $d \geq 1$, no quasi-degeneracy (asymptotic degeneracy in the
thermodynamic limit) can be defined from the energy spectrum.
Exact ground-state degeneracy can still be defined for $d=0$.
However, the exact degeneracy is generally a direct consequence of
Schur's lemma.
Namely, an $SU(N)$ spin in any nontrivial \emph{irreducible} representation
must have exact degeneracy for a $SU(N)$-symmetric Hamiltonian. }

{
Here we consider the spectrum of Hamiltonian imposing only the discrete
subgroup $\mathbb{Z}_N\times\mathbb{Z}_N$ of $PSU(N)$, instead
of the full $PSU(N)$ symmetry.
In this case, an irreducible representation of $PSU(N)$ can become
reducible and the degeneracy of the spectrum may be lifted.
Whether the ground state can be made unique by
such a level splitting is a question analogous to
the ``ingappability'' discussed for $d \geq 1$ dimensions
in the context of the LSM theorem.}

Before the general discussion for $SU(N)$ case, we give several explicit examples for $N=2,3$ cases. If $N=2$, the $\mathbb{Z}_2\times\mathbb{Z}_2$ is generated by $\exp(i\pi S_x/2)$ and $\exp(i\pi S_z/2)$, which, in the fundamental representation $\rho_\text{f}$, satisfy
\begin{eqnarray}
\label{commu_2}
&&\rho_\text{f}[\exp(i\pi S_x/2)]\rho_\text{f}[\exp(i\pi S_z/2)]\nonumber\\
&&\cdot\rho_\text{f}^{-1}[\exp(i\pi S_x/2)]\rho_\text{f}^{-1}[\exp(i\pi S_z/2)]=-1.
\end{eqnarray}
If $N=3$, $\mathbb{Z}_3\times\mathbb{Z}_3$ is generated by $\exp[i\pi (t_8-t_3)/3]$ and $\exp[-i2\pi(t_2-t_5+t_7)/(3\sqrt{3})]$ where, in the fundamental representation, $\rho_\text{f}[t_8]=\text{diag}[1,1,-2]$, $\rho_\text{f}[t_3]=\text{diag}[1,-1,0]$, $\rho_\text{f}[t_2]_{lm}=i\delta_{l,2}\delta_{m,1}-i\delta_{l,1}\delta{m,2}$, $\rho_\text{f}[t_5]_{lm}=i\delta_{l,3}\delta_{m,1}-i\delta_{l,1}\delta_{m,3}$ and $\rho_\text{f}[t_7]_{lm}=i\delta_{l,3}\delta_{m,2}-i\delta_{l,2}\delta_{m,3}$, and, by~\cite{Suppl},
\begin{eqnarray}
\label{commu_3}
&&\rho_\text{f}\left[\exp\left[i\pi\frac{t_8-t_3}{3}\right]\right]\rho_\text{f}\left[\exp\left[-i2\pi\frac{t_2-t_5+t_7}{3\sqrt{3}})\right]\right]\cdot\nonumber\\
&&\cdot\rho_\text{f}^{-1}\left[\exp\left[i\pi\frac{t_8-t_3}{3}\right]\right]\rho_\text{f}^{-1}\left[\exp\left[-i2\pi\frac{t_2-t_5+t_7}{3\sqrt{3}}\right]\right]\nonumber\\
&=&\exp(i2\pi/3).
\end{eqnarray}
The same commutators in Eqs.~(\ref{commu_2},\ref{commu_3}), for the adjoint representation, are trivial~\cite{Suppl} and the constructions of the $\mathbb{Z}_N\times\mathbb{Z}_N$ generators for other $SU(N)$ cases can be found in~\cite{Tanimoto:2015aa}. A general representation with $b$ Young-tableaux boxes for $SU(N)$ cases yields a commutator as $\exp(i2\pi b/N)$, which reflects the $H^2(\mathbb{Z}_N\times\mathbb{Z}_N,U(1))\cong\mathbb{Z}_N$ classification of the projective representation of $\mathbb{Z}_N\times\mathbb{Z}_N$, where $H^2(\cdot,U(1))$ denotes the second group cohomology with a $U(1)$ coefficient ring. Moreover, two transformations with a nontrivial commutator as Eqs.~(\ref{commu_2},\ref{commu_3}) cannot share a single one-dimensional invariant sub-Hilbert-space. These results exactly imply that any $b$ Young-tableaux representation $SU(N)$ ``spin'' can be gapped with a unique ground state respecting $\mathbb{Z}_N\times\mathbb{Z}_N$ \emph{if and only if} $N$ divides $b$.

\begin{figure}
\begin{center}
\includegraphics[width=6.5cm,pagebox=cropbox,clip]{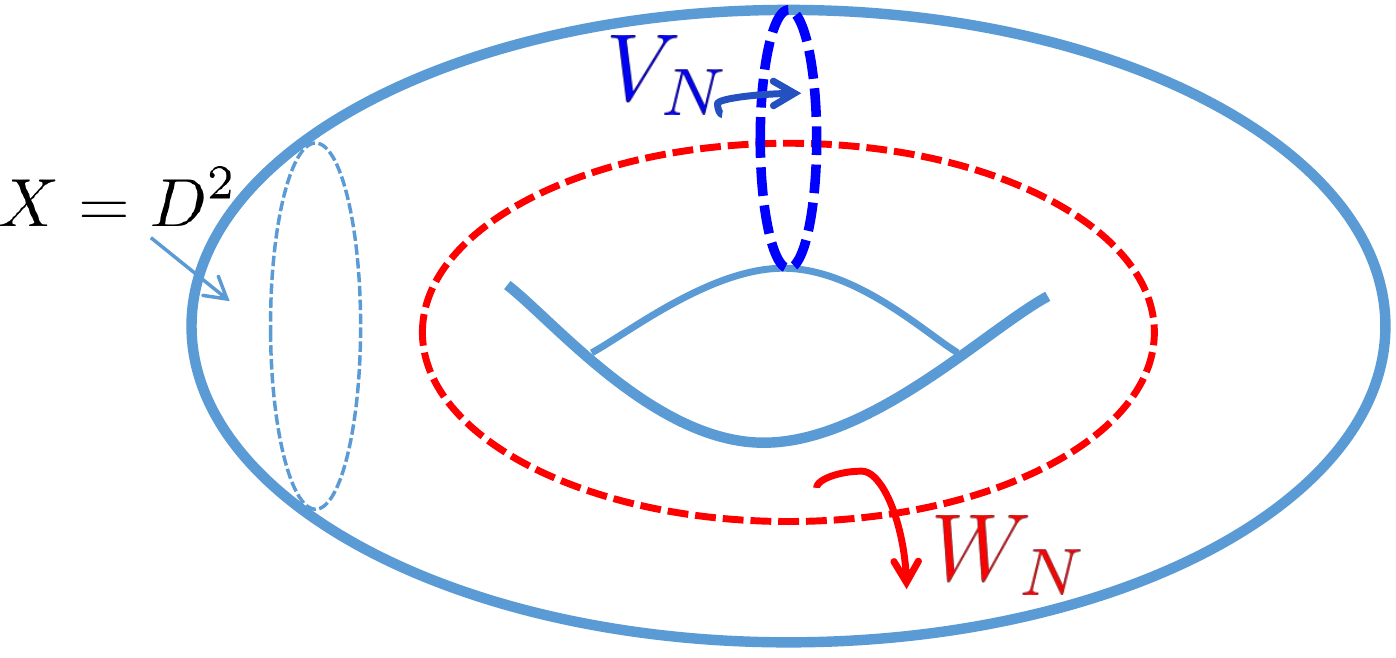}
\caption{$\mathbb{Z}_N\times\mathbb{Z}_N$ bundle on $\tilde{X}=T^2$ where the domain walls are inserted beyond $\partial X=S^1$ and a corresponding transformation is acted across a domain wall according to the domain-wall orientation indicated above. }
\label{torus_0_d}
\end{center}
\end{figure}

Let us discuss and derive such $(0+1)$-dimensional (in)gappabilities in the framework of anomaly.
{The partition function would have a phase ambiguity from the projective representation by an $SU(N)$ spin.
Indeed, the ungauged Dijkgraaf-Witten theory provides a non-Lagrangian construction to expose this relation between the projective representation and $(0+1)$-dimensional anomaly for finite symmetries~\cite{Dijkgraaf:1990aa}.
As follows, we will also manifest this relation in our concrete fermionic representation of spins.

For general $SU(N)$ cases, we first study the fundamental representation case or $b=1$ at a single point.
The low-energy effective theory for the (fine-tuned) lattice Hamiltonian $H_{SU(N)}=0$ is simply:
\begin{eqnarray}
\label{0d}
L_{SU(N)}[{\mathcal A}]&=&\sum_{\alpha=1}^N\bar{\psi}^\alpha(\tau)i(\partial_\tau-i{\mathcal A}_\tau)\psi_{\alpha}(\tau), \end{eqnarray}
defined on the temporal circle $\tau\in S^1=\mathbb{R}/(2\pi\mathbb{Z})$ around which
$\psi(\tau+2\pi)=-\psi(\tau)$, where the minus sign results from the fermionic path integral.
Hence, we do not have $\mathbb{Z}_N\times\mathbb{Z}_N$ twisting around $S^1$.
$\{\psi_\alpha\}_{\alpha=1}^N$ constitutes the projective representation of $\mathbb{Z}_N\times\mathbb{Z}_N$ whose generators $V_N$ and $W_N$ represented by matrices $\rho(V_N)$ and $\rho(W_N)$ satisfying
\begin{eqnarray}
\label{proj}
\rho(V_N)\rho(W_N)\rho^{-1}(V_N)\rho^{-1}(W_N)=\exp(i2\pi/N).
\end{eqnarray}

The $U(1)_Q$ dynamical gauge field ${\mathcal A}_\tau$ is again used to project out the unphysical Hilbert subspace, and the $U(1)_Q$ charge at the site is restricted to be $1$ the same as Eq.~(\ref{restriction}).
Due to this charge number restriction, the response theory of (\ref{0d}) to the gauge field ${\mathcal A}_\tau$ takes the form as
\begin{eqnarray}
\label{z_su(N)}
z_{SU(N)}[{\mathcal A}]&=&|z_{SU(N)}[{\mathcal A}]|\exp\left(-i\int_{S^1}{\mathcal A}_\tau\right),
\end{eqnarray}
by the minimal coupling argument.
The partition function is obtained after the dynamical ${\mathcal A}$ is integrated out:
\begin{eqnarray}
z_{SU(N)}&=&\int\mathscr{D}{\mathcal A}z_{SU(N)}[{\mathcal A}].
\end{eqnarray}
Since the anomaly can be detected by different bulk extensions over which the spin structure and gauge fields extend.
We rewrite $z_{SU(N)}[{\mathcal A}]$ into two different bulk extensions:
\begin{eqnarray}
z_{SU(N)}[{\mathcal A},X]&=&|z_{SU(N)}[{\mathcal A}]|\exp\left(-i\int_{X}{\mathcal F}\right),\\
z_{SU(N)}[{\mathcal A},X']&=&|z_{SU(N)}[{\mathcal A}]|\exp\left(-i\int_{X'}{\mathcal F}\right),
\end{eqnarray}
where $\partial X=\partial X'=S^1$ and they are indeed the extension of $z_{SU(N)}[{\mathcal A}]$ in (\ref{z_su(N)}) since by the Stokes' theorem
\begin{eqnarray}
\int_{X}{\mathcal F}=\int_{X'}{\mathcal F}=\int_{S^1}{\mathcal A}_\tau
\end{eqnarray}
in the absence of $\mathbb{Z}_N\times\mathbb{Z}_N$ background gauge field.
The absolute value of the partition function is well-defined so it does not depend on the extension~\cite{Witten:2016aa}.
$z_{SU(N)}[X]$ and $z_{SU(N)}[X']$ are still defined by integrating out ${\mathcal A}$.
Although we \ed{do not} include any $\mathbb{Z}_N\times\mathbb{Z}_N$ twisting along the physical dimension $S^1$, we can still insert $\mathbb{Z}_N\times\mathbb{Z}_N$ domain walls into the gauge bundle on $X$ and $X'$, which does not affect the trivial $\mathbb{Z}_N\times\mathbb{Z}_N$ bundle on their edge circle $\partial X=\partial X'=S^1$ as follows.
Due to this anti-periodic boundary condition on $S^1$, we can extend it onto the disk $X=D^2$~(See e.g. Chapter~3 of \cite{Bar:2000aa}) and onto $X'$ such that $\tilde{X}\equiv X\cup(-X')=T^2$.
Then the bulk dependence $z_{SU(N)}[{\mathcal A},X]/z_{SU(N)}[{\mathcal A},X']=\exp\left(-i\int_{\tilde{X}}{\mathcal F}\right)$.

The $\mathbb{Z}_N\times\mathbb{Z}_N$ bundle on $\tilde{X}=T^2$ is sketched in FIG.~(\ref{torus_0_d}) where along one cycle of $T^2$ there is a $V_N$ twisting and along the other cycle a $W_N$ twisting. }
To obtain a consistent gauge bundle for the fermionic degrees of freedom on $T^2$, we need to insert a $U(1)_Q$ flux:
\begin{eqnarray}
\int_{\tilde{X}}{\mathcal F}=-\frac{2\pi}{N}\mod 2\pi,
\end{eqnarray}
to induce an Aharonov-Bohm phase exactly eliminating the phase ambiguity brought by the commutator (\ref{proj}).
Thus the phase ambiguity of $z_{SU(N)}[{\mathcal A}]$ due to the different extension $X$ and $X'$ is $\exp(i2\pi/N)$. Furthermore, since such a phase is independent on the fluctuating gauge field ${\mathcal A}$ which is integrated within one gauge sector connected by small gauge transformations, we can extract this phase out of the integration $\int\mathscr{D}{\mathcal A}$. Thus this phase ambiguity denoted by $Z_{SU(N)}$ is also shared by $z_{SU(N)}$ after such a functional integration:
\begin{eqnarray}
Z_{SU(N)}\equiv\frac{z_{SU(N)}[X]}{z_{SU(N)}[X']}=\exp(i2\pi/N).
\end{eqnarray}
By additivity of anomaly, we have for general representations with total $b$ of Young-tableaux boxes:
\begin{eqnarray}
\label{ano_su(N)_0d}
Z^{(b)}_{SU(N)}&=&\left(Z_{SU(N)}\right)^b=\exp\left( i 2\pi\frac{b}{N}\right).
\end{eqnarray}
This precisely implies that the partition function is enjoying an anomaly resultant from a projective representation of $\mathbb{Z}_N\times\mathbb{Z}_N$.
Indeed, when $b$ is divisible by $N$, such as spin-$1$ (adjoint) representation of $su(2)$, the interaction $S_z^2$ can gap the ground state uniquely preserving $\mathbb{Z}_2\times\mathbb{Z}_2$ generated by $\exp(i\pi S_x/2)$ and $\exp(i\pi S_z/2)$. For the adjoint representation of $su(3)$, which is made of three Young-tableaux boxes, the ground state can be also gapped uniquely since the generators of $\mathbb{Z}_N\times\mathbb{Z}_N$ share at least one, actually eight, one-dimensional invariant sub-Hilbert-spaces~\cite{Suppl}.

Including the exact degeneracies in $d\geq1$ when $bV/N\notin\mathbb{N}$ as well, Eq.~(\ref{ano_su(N)}) together with Eq.~(\ref{ano_su(N)_0d}) exactly gives the following generalized LSM theorem in any dimension of $d\geq0$, which has a well-defined thermodynamic limit independent on the system size specification:

\emph{If a $(d>0)$-dimensional Hamiltonian possesses $SU(N)$ spin-rotation and translation symmetries, or if a $(d=0)$-dimensional Hamiltonian possesses $\mathbb{Z}_N\times\mathbb{Z}_N$, the system does not permit a trivially gapped phase when the total number $b$ of Young-tableaux boxes per unit cell is not divisible by $N$. The converse is true if $d=0$. }

It is an appropriate point to remark that the role played by filling constraint is different between $d=0$ and $d>0$. In the former case, the filling condition restricts the theta term of a possible bulk extension, while filling restricts the form of translation symmetry at the low-energy continuum limit in $d>0$.
Moreover, unlike LSM theorem in arbitrary dimensions discussed in Sec.~\ref{u(1)_LSM}, our generalization of LSM does not have a compact form when $d=0$ is included, because of the higher symmetry $PSU(N)$ than $U(1)_Q$.

\subsection{Bulk-boundary correspondences: LSM theorem with $PSU(N)$}
\label{bulk-boundary_psu(N)}
{In this Subsection, we will construct the SPT ``bulk'' theory with $PSU(N)$
symmetry
in one higher dimensions, which has the boundary theory with the LSM-type anomaly, in two different ways.}

\subsubsection{A weak SPT point of view}
To check whether the lattice system has any other LSM-type anomaly beyond Eq.~(\ref{ano_su(N)}), we make use of a weak-SPT viewpoint on quantum anomaly calculated before. Since the quantum anomaly of a spatially $d$-dimensional system can be understood as that of the boundary of a $((d+1)+1)$-dimensional bulk SPT phase. Therefore, the relevant SPT bulk that the LSM-type anomaly of system corresponds to is assumed to lie in the classes of $H^{d+2}(PSU(N)\times\mathbb{Z}^d,U(1))$ which is the $PSU(N)$ (group) cohomology group with $U(1)$ as a (trivial) group-module~\cite{Gu:2009aa}, where $\mathbb{Z}^d$ is a direct product of $d$ of lattice translations corresponding to the translations of its $d$-dimensional spatial boundary in the thermodynamical limit~\cite{Cheng:2016aa,Xiong:2018aa}. By K{\"u}nneth formula,
\begin{eqnarray}
\label{kunneth_0}
&&H^{d+2}(PSU(N)\times\mathbb{Z}^d)\nonumber\\
&=&H^{d+2}(PSU(N)\times\mathbb{Z}^{d-1})\oplus H^{d+1}(PSU(N)\times\mathbb{Z}^{d-1})\nonumber\\
&=&\cdots\nonumber\\
&=&\left[\oplus_{k=3}^{d+2}H^{k}(PSU(N)\times\mathbb{Z}^{k-3})\right]\oplus H^2(PSU(N)),
\end{eqnarray}
where, for clearness, the coefficient ring $U(1)$ is suppressed. Let us explain the first line of the equation above. $H^{d+2}(PSU(N)\times\mathbb{Z}^{d-1})$ represents the $((d+1)+1)$-dimensional SPT phases protected by $PSU(N)\times\mathbb{Z}^{d-1}$ where the $d$-th $\mathbb{Z}$ is neglected, {which is consistent with the projection $H^{d+2}(PSU(N)\times\mathbb{Z}^{d})\rightarrow H^{d+2}(PSU(N)\times\mathbb{Z}^{d-1})$ where the group structure of $d$-th $\mathbb{Z}$ is forgotten.}
$H^{d+1}(PSU(N)\times\mathbb{Z}^{d-1})$ corresponds to the $(d+1)$-dimensional SPT phases constructed by stacking, uniformly in the $d$-th direction, $d$-dimensional SPT's protected by $PSU(N)\times\mathbb{Z}^{d-1}$.
{This correspondence can be derived in a nonlinear sigma model (with target space as classifying spaces of groups~\cite{Chen:2013aa}) approach~(See e.g. Appendix~C. in \cite{Xiong:2018aa}). Intuitively speaking, the $d$-th translation $\mathbb{Z}$ introduces an obstruction for the phase to be trivialized with unit cells along that direction.}

{The stacking constructions of SPT classes in $H^{d+2}(PSU(N)\times\mathbb{Z}^d)$ can be also justified after we continue to apply the K{\"u}nneth formula to (\ref{kunneth_0}) to obtain
\begin{eqnarray}
H^{d+2}(PSU(N)\times\mathbb{Z}^d)=\oplus_{r=0}^{d}\left[H^{r+2}(PSU(N))\right]^{\frac{d!}{(d-r)!r!}}, \nonumber\\
\end{eqnarray}
where the copy number $d!/[(d-r)!r!]$ of $H^{r+2}(PSU(N))$ is exactly the total number of perpendicular directions to stack $((r+1)+1)$-dimensional $PSU(N)$-SPT phase by the translations $\mathbb{Z}^d$.}
Therefore, the last term $H^2(PSU(N))$ means $((d+1)+1)$-dimensional SPT phases stacked by copies of a one-dimensional SPT phase (or chain) protected by $PSU(N)$ uniformly in all the $d$ spatial dimensions while all the $d$ of translations are obstruction for it to be trivialized.
One {characterizing} property of phases by $H^2(PSU(N))$ is that lifting any of the $d$ translations and $PSU(N)$ out will trivialize the phase, while other components do not have this property.
In this sense, $H^2(PSU(N))$ is the maximally mixed anomaly of $PSU(N)\times\mathbb{Z}^d$.

On the other hand of our current system, the generalized LSM also has the same ``mixing'' characteristic since any $N$ of unit cells can be $SU(N)$ singlet if any of the translations are permitted to be broken explicitly, and $PSU(N)$ symmetry is obviously essential for the ingappability as well.
Thus the LSM-type anomaly of our system is a class in $H^2(PSU(N))$.
Conversely, the anomaly factor in Eq.~(\ref{ano_su(N)}) exactly implies a $\mathbb{Z}_N$ group structure and the generator is the fundamental case: $b=1$. Therefore, $b=1,\cdots,N$ represents the $\mathbb{Z}_N$ elements. Since $H^2(PSU(N))\cong\mathbb{Z}_N$ thereby saturated by $b=1,\cdots,N$, we can come to the conclusion that all possible LSM-type anomalies have been extracted out by Eq.~(\ref{ano_su(N)}) with $b=1,2,\cdots,N$, which are exactly classified by $H^2(PSU(N))$.

\subsubsection{A BF-theory approach}
Similarly to the $U(1)_Q$ case studied in details in Sec.~\ref{bulk-boundary_u(1)}, we can also construct the bulk massive fermions with opposite masses for the two opposite chiralities of Dirac fermions and derive the response theory for the fundamental chain $(b=1)$ as the following BF theory by replacing $A_{U(1)}$ in (\ref{bf_u(1)}) by {$A_{PSU(N)}+{\mathcal A}\equiv A_{U(N)}+{\mathcal A}'$ with a dynamical $U(1)$-gauge field ${\mathcal A}'$ properly quantized as $\int d{\mathcal A}'\in2\pi\mathbb{Z}$}:
\begin{eqnarray}
&&Z^{(b=1)}_{\text{bulk},PSU(N)}\nonumber\\
&=&\int\mathscr{D}{\mathcal A}\exp\left\{\int\frac{i}{2\pi}\text{Tr}
\left[a \wedge d(A_{PSU(N)}+\mathcal A)\right]\right\}\nonumber\\
&=&{\int\mathscr{D}{\mathcal A}'\exp\left\{\int\frac{i}{2\pi}\text{Tr}
\left[a \wedge d(A_{U(N)}+{\mathcal A}')\right]\right\}}\nonumber\\
&=&\exp\left\{\int\frac{i}{2\pi}\text{Tr}
\left[a \wedge dA_{U(N)}\right]\right\},
\end{eqnarray}
where the smooth sector of the dynamical field ${\mathcal A}'$ plays the role as a Lagrangian multiplier and {its topologically nontrivial sectors} restrict the translation-gauge field $a$ to satisfy (See e.g. Appendix C.3. of~\cite{Seiberg:2016aa}):
\begin{eqnarray}
\label{z_n_constraint}
da=0\,\,\text{ and }\oint_\text{closed loop}a\in\frac{2\pi}{N}\mathbb{Z}.
\end{eqnarray}
Comparing (\ref{z_n_constraint}) with (\ref{z_q_constraint}) justifies the effective filling factor (\ref{effective_filling}).
We can take the following gauge bundle {on the mapping torus} $T^2\times S^1$:
\begin{eqnarray}
\label{background_bulk}
&&\int_{S^1}a=\frac{2\pi}{N} \,\,\text{ and}\,\,\int_{T^2}\text{Tr}\left[\frac{dA_{U(N)}}{2\pi}\right]=1,
\end{eqnarray}
where $a$ is a pull-back from a flat gauge field on the extra dimension $S^1$ and $A_{U(N)}$ is a pull-back from a gauge field on the physical spacetime $T^2$ representing a flux insertion.
{The translation-symmetry holonomy $\int_{S^1}a$ in (\ref{background_bulk}) realizes the translation-symmetry twisting of bulk field and thus mapping-torus partition function $Z^{(b=1)}_{\text{bulk},PSU(N)}$ exactly reproduces the phase ambiguity (\ref{partition_su(N)}) brought by a translation transformation. }

Therefore, due to the additivity of the bulk,
\begin{eqnarray}
Z^{(b)}_{\text{bulk},PSU(N)}&=&\left(Z^{(b=1)}_{\text{bulk},PSU(N)}\right)^b\nonumber\\
&=&\exp\left(i2\pi\frac{b}{N}\right),
\end{eqnarray}
which is the same anomaly factor as Eq.~(\ref{ano_su(N)_0d}).

\section{TLBC applied to the ingappabilities by time reversal}
\label{time_reversal}
In this Section, we will discuss the application of TLBC to the
ingappability constrained by a time reversal symmetry.
As an example, let us consider the following half-filled $N$-flavor spinless fermion on a square lattice with $\pi$-flux per plaquette:
\begin{eqnarray}
H_\pi&=&\sum_{f=1}^Nt\left(\sum_{\vec{r}}c^{\dagger}_{(\vec{r}+\hat{x})f}c_{\vec{r}f}+\text{h.c.}\right)\nonumber\\
&&+\sum_{f=1}^Nt\left[\sum_{\vec{r}}c^{\dagger}_{(\vec{r}+\hat{y})f}c_{\vec{r}f}(-1)^{r_1}+\text{h.c.}\right]\nonumber\\
&\equiv&H_{\pi1}+H_{\pi2},
\label{eq.Hpi}
\end{eqnarray}
where we have chosen a gauge such that $H_\pi$ is still lattice-translation symmetric along $\hat{y}$. Nevertheless, the translation symmetry along $\hat{x}$ seems to be broken to two sites. However, physically, the system should be still symmetric along $\hat{x}$ with one site, and such a translation symmetry and the original translation along $\hat{y}$ in the current gauge choice, called magnetic translations~\cite{Zak:1964aa,Zak:1964ab} equally for all flavors $f$'s satisfies:
\begin{eqnarray}
\label{gauge_fix}
T_1c_{\vec{r}f}T_1^{-1}&=&c_{(\vec{r}+\hat{x})f}\exp( i \pi r_2), \nonumber\\
T_2c_{\vec{r}f}T_2^{-1}&=&c_{(\vec{r}+\hat{y})f}.
\end{eqnarray}
We will use $T_{1,2}$ to denote the magnetic translations in the following discussion, instead of original lattice translations.
To make $T_1$ a well-defined unitary transformation representing an exact symmetry, we need to impose that $L_y\in2\mathbb{N}$.
Then the charge quantization in Eq.~(\ref{charge_quantization}) is automatically fullfilled.
It should be noted that the gauge-invariant nature of $T_{1,2}$ is encoded in their commutator:
\begin{eqnarray}
\label{commutator_inv}
T_1T_2T_1^{-1}T_2^{-1}=-1.
\end{eqnarray}
Here ``half-filled'' implies that the particle number per physical $1\times1$ unit cell is $1/2$.
However, the LSM theorem, even after generalized to $U(N)$ cases, cannot say anything nontrivial for the present case with the magnetic translation symmetry.
Indeed, the following interaction, respecting $U(N)$ and $T_{1,2}$, can open a gap with a unique ground state:
\begin{eqnarray}
\label{dH}
\Delta H&=&\sum_{\vec{r},f}t(-1)^{r_1}\left[ i c_{\vec{r}f}^{\dagger}c_{(\vec{r}+\hat{x}+\hat{y})f}+c_{\vec{r}f}^{\dagger}c_{(\vec{r}+3\hat{x}+\hat{y})f}\right]+\text{h.c.}. \nonumber\\
\end{eqnarray}
It should be noted, however, that $\Delta H$ breaks the time-reversal symmetry explicitly.
Thus we may expect that imposing time-reversal symmetry in addition to the
magnetic translation symmetry potentially obstructs the trivially gapping.

Indeed, it has been proposed that, in $N=1$ case, the Hall conductance of the system when trivially gapped by a $U(1)_Q$ and $T_{1,2}$ symmetric interaction, must be odd~\cite{Dana:1985aa,KolRead1993,Lu:2017aa,Matsugatani:2018}.
This implies that, there cannot be a unique ground state with a non-vanishing excitation gap, when the time-reversal symmetry is additionally imposed.
However, the arguments in Refs.~\cite{Dana:1985aa,KolRead1993,Lu:2017aa,Matsugatani:2018} again rely on the special choice of the system sizes.
Here we apply the TLBC to the systems invariant under
the magnetic translation and the time reversal, which
reveals an anomaly manifestation of the ingappability constraint
due to the time-reversal symmetry in the context of field theory.

\subsection{Low-energy effective theory and symmetries}

Our strategy is to apply the TLBC to the low-energy effective field theory.
As a preparation, we first use the PBC to define the momentum
representation (Fourier components)
\begin{eqnarray}
c_{(2n_1,r_2)}&=&\sum_{\vec{k}}a_{\vec{k}}\exp\left( i k_1n_1+ i k_2r_2\right),\nonumber\\
c_{(2n_1+1,r_2)}&=&\sum_{\vec{k}}b_{\vec{k}}\exp\left( i k_1n_1+ i k_2r_2\right),
\end{eqnarray}
where $\vec{k}\in(-\pi,\pi]\times(-\pi,\pi]$.
We will introduce the TLBC in the next subsection.
The momentum representation of the tight-binding Hamiltonian~\eqref{eq.Hpi}
reads
\begin{eqnarray}
H_{\pi1}&=&t\sum a^\dagger_kb_k[1+\exp(- i k_1)]+b^\dagger_ka_k[1+\exp( i k_1)],\nonumber\\
H_{\pi2}&=&t\sum 2\left(a^\dagger_ka_k-b^\dagger_kb_k\right)\cos k_2.
\end{eqnarray}
Therefore,
\begin{eqnarray}
H_\pi=t\sum (a^\dagger_k,b^\dagger_k)\!\left(\begin{array}{cc}2\cos k_2&1+\exp(- i k_1)\\1+\exp( i k_1)&-2\cos k_2\end{array}\right)\!\left(\!\!\begin{array}{c}a_k\\b_k\end{array}\!\!\!\right). \nonumber
\end{eqnarray}
The low-energy excitation momentum points are localized around
\begin{eqnarray}
K&=&\left(\pi,-\pi/2\right)\text{ and }K'=\left(\pi,+\pi/2\right).
\end{eqnarray}
The low-energy effective theory can be obtained as,
\begin{eqnarray}
H_\pi\approx t\sum_{l=\{1,2\};k}\psi_{k}^{\dagger(l)}(-2\sigma_3k_2-\sigma_2k_1)\psi^{(l)}_{k},
\end{eqnarray}
where
\begin{eqnarray}
\psi_{k}^{(1)}&=&\sigma_2\left(\begin{array}{c}a_{k+K}\\b_{k+K}\end{array}\right)=\left(\begin{array}{c}- i b_{k+K}\\ i a_{k+K}\end{array}\right),\\
\psi_{k}^{(2)}&=&\left(\begin{array}{c}a_{k+K'}\\b_{k+K'}\end{array}\right).
\end{eqnarray}
We have the following symmetry representation:
\begin{itemize}
\item{{Flavor $U(N)$ symmetry $U_{\{\phi\}}$}

Conventionally, the global $U(N)$ symmetry is defined as
\begin{eqnarray}
U_{\{\phi\}} c_{\vec{r}}U^{-1}_{\{\phi\}}=\exp\left(\sum_{k=0}^{N^2-1}-i\phi_kt_k\right)c_{\vec{r}},
\end{eqnarray}
where $t_k$'s: $t^{ff'}_0=\delta_{f,f'}$ the $U(1)$ generator while $\{t_k:k=1,2,\cdots,N^2-1\}$ are $SU(N)$ generators in the fundamental representation with a renormalization such that $\phi_k$'s are each compactified by $2\pi$.
Equivalently,
\begin{eqnarray}
U_{\{\phi\}}=\exp\left(\sum_{\vec{r},f,f'}\sum_{k=0}^{N^2-1}ic^{\dagger}_{\vec{r}f}t_k^{ff'}\phi_kc_{\vec{r}f'}\right).
\end{eqnarray}
Since $U(1)$ and $SU(N)$ share a center, the definition above is not faithful and a more systematic approach is to impose a global structure both on $U(1)$ and $SU(N)$ parameters $\{\phi_k\}$ by a quotient over $\mathbb{Z}_N$ since $U(N)\cong[U(1)\times SU(N)]/\mathbb{Z}_N$.

For the low-energy degrees of freedom,
\begin{eqnarray}
U_{\{\phi\}}\psi U^{-1}_{\{\phi\}}=\exp\left(\sum_{k=0 }^{N^2-1}-i\phi_kt_k\right)\psi.
\end{eqnarray}
In the following discussion, the flavor indices ``$f$'' will be suppressed for simplicity.
}
\item{{Magnetic translation $T_1$}

The lattice Hamiltonian is invariant under the magnetic translation along $\hat{x}$-axis:
\begin{eqnarray}
T_1c_{\vec{r}}T_1^{-1}=c_{\vec{r}+\hat{x}}\exp( i \pi r_2),
\end{eqnarray}
which means
\begin{eqnarray}
T_1a_kT_1^{-1}&=&b_{k+Q},\\
T_1b_kT_1^{-1}&=&a_{k+Q}\exp( i k_1),
\end{eqnarray}
{where $Q\equiv K'-K=(0,\pi)$. }

In the low-energy field theory,
\begin{eqnarray}
T_1\psi T_1^{-1}= i \tau_1\psi,
\end{eqnarray}
where $\tau$ matrices act on the valley components.
}
\item{{Translation $T_2$}

The Hamiltonian is also invariant under the conventional translation along $\hat{y}$-axis:
\begin{eqnarray}
T_2c_{\vec{r}}T_2^{-1}=c_{\vec{r}+\hat{y}},
\end{eqnarray}
which gives
\begin{eqnarray}
T_2a_kT_2^{-1}&=&a_k\exp( i k_2),\\
T_2b_kT_2^{-1}&=&b_k\exp( i k_2),
\end{eqnarray}
and
\begin{eqnarray}
T_2\psi T_2^{-1}=- i \tau_3\psi.
\end{eqnarray}
}
\item{{Time-reversal symmetry $\mathbb{Z}_2^T$}

Since we are interested in the constraint brought by time-reversal symmetry $\mathbb{Z}_2^T$, we define the following (anti-unitary) time-reversal symmetry on our spinless fermions as
\begin{eqnarray}
\Theta_0 i \Theta_0^{-1}=- i ,\\
\Theta_0c_{\vec{r}}\Theta_0^{-1}=c_{\vec{r}}.
\end{eqnarray}
Thus
\begin{eqnarray}
\Theta_0a_k\Theta_0^{-1}&=&a_{-k},\\
\Theta_0b_k\Theta_0^{-1}&=&b_{-k},
\end{eqnarray}
and
\begin{eqnarray}
\Theta_0\psi_{\vec{k}}\Theta_0^{-1}=- i \tau_2\otimes\sigma_2\psi_{-\vec{k}},
\end{eqnarray}
or
\begin{eqnarray}
\Theta_0\psi_{1,k}\Theta_0^{-1}&=&-\sigma_2\psi_{2,-k},\nonumber\\
\Theta_0\psi_{2,k}\Theta_0^{-1}&=&\sigma_2\psi_{1,-k}.
\end{eqnarray}
}
\end{itemize}
Let us do a trivial rescaling $k_2\rightarrow k_2/2$ and $t\rightarrow1$. Back to the real space, the Hamiltonian density becomes:
\begin{eqnarray}
\label{critical}
\mathscr{H}&=&\sum_{f=1}^{N}\sum_{l=1}^2 i \psi_{f}^{\dagger(l)}(\sigma_3\partial_2+\sigma_2\partial_1)\psi^{(l)}_{f}\nonumber\\
&=&- i \bar{\psi}(-\gamma^0\sigma_2\partial_1-\gamma^0\sigma_3\partial_2)\psi\nonumber\\
&=&- i \bar{\psi}(\gamma^1\partial_1+\gamma^2\partial_2)\psi,
\end{eqnarray}
where, for clearness, we have suppressed the summation over flavor ``$f$'' and valley ``$l$'' indices. $\bar{\psi}\equiv\psi^\dagger\gamma^0$ and $\left\{\gamma^\mu,\gamma^\nu\right\}=2\eta^{\mu\nu}=2\cdot\text{diag}(+,-,-)$, the Dirac algebra in $(1+2)$ dimensions. We can choose the following basis:
\begin{eqnarray}
\left\{\begin{array}{l}\gamma^0=\sigma_1,\\\gamma^1=- i \sigma_3,\\\gamma^2= i \sigma_2.\end{array}\right.
\end{eqnarray}
Then the Lagrangian density is $\mathscr{L}_\text{Dirac}=\bar{\psi} i \gamma^i\partial_i\psi$ with the following symmetry representations:
\begin{eqnarray}
\hat{T}_1\psi(t,x,y)&=& i \tau_1\psi(t,x,y),\nonumber\\
\hat{T}_2\psi(t,x,y)&=&- i \tau_3\psi(t,x,y),\nonumber\\
\hat{\Theta}_0\psi(t,x,y)&=&-\tau_2\otimes\gamma^2\psi^*(-t,x,y)\nonumber\\
&=&-\tau_2(\Theta_\text{E}\psi)(t,x,y)
\end{eqnarray}
where $\Theta_\text{rel}=\gamma^2K$ is the emergent relativistic time-reversal of Dirac spinor with $K$ the antilinear operator $K i =- i K$.
After a Wick rotation $\partial_t\rightarrow i\partial_\tau$, the eigenvalue problem of the corresponding Euclidean Dirac operator reads
\begin{eqnarray}
\label{eom}
\Gamma^i\partial_i\psi(\tau,x,y)=\lambda\psi(\tau,x,y),
\end{eqnarray}
in which $\{\Gamma^j,\Gamma^k\}=-2\delta^{jk}=-2\cdot\text{diag}(+,+,+)$ and
\begin{eqnarray}
\left\{\begin{array}{l}\Gamma^0= i \sigma_1,\\\Gamma^1=- i \sigma_3,\\\Gamma^2= i \sigma_2.\end{array}\right.
\end{eqnarray}

The symmetry field-configuration representations is analytically continued to~\cite{Witten:2016aa}
\begin{eqnarray}
\hat{T}_1\psi(\tau,x,y)&=& i \tau_1\psi(\tau,x,y),\nonumber\\
\hat{T}_2\psi(\tau,x,y)&=&- i \tau_3\psi(\tau,x,y),\nonumber\\
\label{wick_rotation}
\hat{\Theta}_0\psi(\tau,x,y)&=&- i \tau_2\otimes\Gamma^2\Gamma^0\psi^*(-\tau,x,y)\nonumber\\
&=&-\tau_2\otimes (CR_\tau\psi)(\tau,x,y),
\end{eqnarray}
with $C\psi(\tau,x,y)=\Gamma^2\psi^*(\tau,x,y)$ the charge conjugation and $R_\tau\psi(\tau,x,y)= i \Gamma^0\psi(-\tau,x,y)$ the reflection about the temporal direction.
{The analytical continuation $\Theta_\text{rel}\mapsto CR_\tau$ above can be understood from their behaviors on the gauge field~\cite{Witten:2016aa}: $\Theta_\text{rel}A_\mu(t,\vec{x})=(-1)^{1+\delta_{\mu,t}}A_\mu(-t,\vec{x})$, $CA_i(\tau,\vec{x})=(-1)A_i(\tau,\vec{x})$ and $R_\tau A_i(\tau,\vec{x})=(-1)^{\delta_{i,\tau}}A_i(-\tau,\vec{x})$, which implies that $CR_\tau$ reproduces $\Theta_\text{rel}$. }

\subsection{TLBC for the effective field theory}

Let us now impose the TLBC for the low-energy degrees of
freedom $\psi(\tau,x,y)$ as
\begin{eqnarray}
\label{tlbc}
\left\{\begin{array}{l}\psi(\tau,x+L_x,y)=T_2\psi(\tau,x,y); \\
\psi(\tau,x,y+L_y)=\psi(\tau,x,y). \end{array}\right.
\end{eqnarray}
Then we do a flux insertion into the hole rounded by $L_x$:
\begin{eqnarray}
\label{connection}
\left\{\begin{array}{l}A_\tau(\tau,x,y)=0;\\
A_x(\tau,x,y)=\frac{\pi}{L_x}\frac{\tau}{L_\tau}t_0;\\
A^\theta_y(\tau,x,y)=\frac{\theta}{L_y}\frac{1}{N}(t_0-t_{N^2-1}),
\end{array}\right.
\end{eqnarray}
where $t_0$ is an $(N\times N)$ identity matrix and $t_{N^2-1}=\text{diag}[1,1,\cdots,1,-(N-1)]$. For later use, we also formally introduce a static flux into the hole of $L_y$, but we take $\theta=0\mod2\pi$ later, which is gauge equivalent to $A_y=0$ by a large $U(N)$-gauge transformation as $\exp\left[i2\pi y(t_0-t_{N^2-1})/(NL_y)\right]$. Afterwards, we further do a $T_1$ transformation.
Then the Euclidean path-integral partition function related to this flux insertion followed by $T_1$ transformation takes the form as:
\begin{eqnarray}
\label{z}
z(\theta)&\equiv&\int\mathscr{D}(\bar{\psi}_\alpha,\psi_\alpha)\exp\left(-\int\mathscr{L}_\text{Dirac}\left[A_x(\tau),A_y^\theta\right]\right)\nonumber\\
\end{eqnarray}
with the space-time manifold as a four-dimensional torus $T^4$ and $\mathscr{T}$ the time-ordering operator with the boundary condition combined with Eq.~(\ref{tlbc}) as
\begin{eqnarray}
\label{tlbc_tau}
\left\{\begin{array}{l}
\psi(\tau+L_\tau,x,y)=-T_1\psi(\tau,x,y); \\\psi(\tau,x+L_x,y)=T_2\psi(\tau,x,y); \\
\psi(\tau,x,y+L_y)=\psi(\tau,x,y),  \end{array}\right.
\end{eqnarray}
where the minus sign in the boundary condition along $\tau$ is due to the fermionic nature of the path integral.

TLBC by $T_{1,2}$ is obviously consistent with $\Theta_0$ since $[T_{1,2},\Theta_0]=0$.
Nevertheless, one might have noticed that the flux-insertion by $A_x$ is only $\pi$ flux through the $\tau$-$x$ plane rather than $2\pi$. At the first glance, it gives an ill-defined transition function across the cut between  $\tau=L_\tau$ and $\tau=0$. However, it is actually canonical because the latter $T_1$ transformation at $\tau=L_\tau$ anticommutes with $T_2$ imposed in $\hat{x}$ direction: $T_1T_2T_1^{-1}T_2^{-1}=-1$, and it makes the boundary condition Eq.~(\ref{tlbc_tau}) appear inconsistent, either. Such an extra ``$-1$'' sign inconsistency in the boundary condition exactly compensates the transition-function sign ambiguity brought by a $\pi$-flux insertion by $A_x(\tau,x,y)$ in Eq.~(\ref{connection}). A systematic construction of the gauge bundle above is given in~\cite{Suppl}.

Then let us take a look at the following ratio
\begin{eqnarray}
\label{phase_ambiguity_0}
Z_\pi\equiv\frac{z(\theta=2\pi)}{z(\theta=0)},
\end{eqnarray}
{
and the phase of $Z_\pi$ denotes the difference between the momentum transfers in the presence of $A_y^{\theta=2\pi}$ and $A_y^{\theta=0}$ (c.f. Eqs.~(\ref{z},\ref{tlbc_tau})).
Thus, due to the gauge equivalence of $A_y^{\theta=2\pi}$ and $A_y^{\theta=0}$, $Z_\pi$ is $1$ if the theory is anomaly-free.}
Thus the discrepancy between $Z_\pi$ and $1$ signals the gauge anomaly of our Dirac theory (\ref{critical}).

To evaluate this anomaly factor $Z_\pi$, we can first introduce a Pauli-Villars regulator Lagrangian $\sum_{\alpha,f} i \bar{\xi}_{\alpha,f}\left[ i \Gamma^i(\partial_i- i A_i)-m_\text{P-V}\right]\xi_{\alpha,f}$ to regularize the $3$D partition function, where $\xi_\alpha$ is a two-flavor bosonic spinor and the diagonal mass term ($m_\text{P-V}\delta^{\alpha\beta}\delta^{ff'}$) preserves $U(N)$ and $T_{1,2}$.
The partition function is regularized as $|Z_\text{$3$D}|\exp(-i\pi\eta_\text{$3$D}/2)$ explicitly breaking $\mathbb{Z}_2^T$ due to the regulator mass, where $|Z_\text{$3$D}|$ is the absolute value of our massless theory partition function and $\eta_\text{$3$D}\equiv\lim_{\varepsilon\rightarrow0^+}\sum_{k}\text{sgn}(\lambda_k)\exp(-\varepsilon\lambda_k^2)$ with $\{\lambda_k\}$ the spectrum of $3$D two-flavor Dirac operator.
However, $\mathbb{Z}_2^T$ can be restored when we attach a $4$D bulk on it, or equivalently, add a $4$-D counter-term supported by $X':\partial X'=T^3$ whose boundary is our $3$D manifold $T^3$: $2\left[({2}/{48})\int_{X'}{\text{tr}R\wedge R}/{(2\pi)^2}+\int_{X'}({1}/{2}){\text{Tr}F\wedge F}/{(2\pi)^2}\right]$, where $R$ is the $2$-form curvature tensor and ``tr'' is taken over the four spacetime indices and ``Tr'' over the flavors.
Thus the regularized partition function on $X'$ by Atiyah-Patodi-Singer (APS) index theorem~\cite{Atiyah:1975ab} is
\begin{eqnarray}
\label{bulk_edge_correspondence}
Z_\text{$3$D}(X')&=&|Z_\text{$3$D}|\exp\left(-i\pi{\mathcal I}_\text{$4$D}(X')\right), \\
\label{bulk_attachment}
{\mathcal I}_\text{$4$D}(X')&=&\frac{\eta_\text{$3$D}(\partial X')}{2}-\left[\int_{X'}\frac{N}{24}\frac{\text{tr}R\wedge R}{(2\pi)^2}+\frac{\text{Tr}F\wedge F}{(2\pi)^2}\right], \nonumber\\
\end{eqnarray}
in which $Z_\text{$3$D}(X')$ explicitly respects time-reversal symmetry since ${\mathcal I}_\text{$4$D}$ of a $4$-D Dirac operator with $N$ flavors and $2$ valleys, the number difference of zero modes between positive and negative chiralities, is proportional to the effective action of a $(3+1)$-dimensional massive fermion and it is an integer calculated under APS boundary condition.
$Z_\text{$3$D}(X')$ potentially depends on the extension $X'$, which disables our massless system to be purely a $3$D model.
\begin{figure}
\begin{center}
\includegraphics[width=6.cm,pagebox=cropbox,clip]{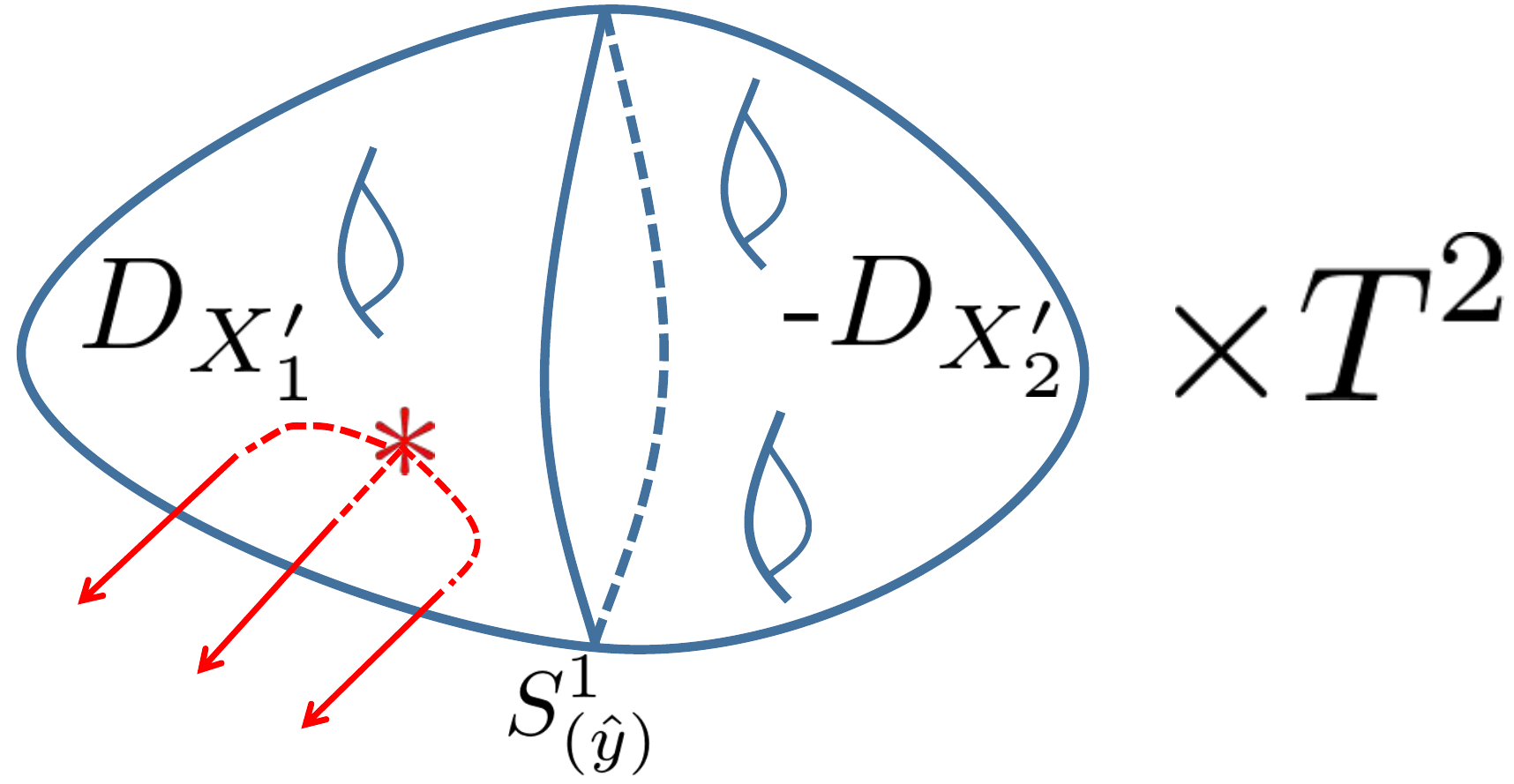}
\caption{$X'_{1,2}$ are glued along their common boundary $T^3$. There is a unit $U(N)$ instanton labelled by ``{\color{red}\textasteriskcentered}'' inside $D_{X_1'}\cup-D_{X_2'}$ inducing a flux through $D_{X_1'}$. }
\label{glued}
\end{center}
\end{figure}
{As shown below, $Z_\pi$ in (\ref{phase_ambiguity_0}) can be calculated by a phase ambiguity induced by two different extensions, with and without, respectively, a $2\pi(t_0-t_{N^2-1})/N$-flux in the orientable $D_{X'}$ spanned by the extra dimension and the $\hat{y}$-direction circle $S^1_{(\hat{y})}$ (non-orientable extensions to be discussed later).
Thus $X'=T^2\times D_{X'}$ and we also set $\partial D_{X'}=S^1_{(\hat{y})}$ so that $\partial X'=T^3$ is the spacetime of our $(2+1)$-dimensional system.
We take $X'_{1,2}=T^2\times D_{X'_{1,2}}$ to extend both the space-time and the gauge field, where $D_{X'_1}$ and $D_{X'_2}$ are not necessarily the same.
We insert $2\pi(t_0-t_{N^2-1})/N$-flux through $D_{X_1'}$ as in FIG.~(\ref{glued}) and we obtain $A_y=A_y^{\theta=2\pi}$ on its boundary $S^1_{(\hat{y})}$ by the Stokes' theorem up to a small gauge transformation.
On the other side, $D_{X'_2}$ has no flux through it.
Then $A_y=A_y^{\theta=0}$ on its boundary $S^1_{(\hat{y})}$ up to a small gauge transformation.
Therefore, the bulk-extension dependence here precisely characterizes the gauge anomaly (\ref{phase_ambiguity_0}) of the boundary theory between the two different gauge choices $A_y^{\theta=2\pi}$ and $A_y^{\theta=0}$:
\begin{eqnarray}
\label{phase_ambiguity}
Z_\pi&=&\frac{Z_\text{3D}(X'_1)}{Z_\text{3D}(X'_2)}=(-1)^{{\mathcal I}_\text{$4$D}(X'_1\cup-X'_2)}=-1,
\end{eqnarray}
where we have used the index pasting rule ${\mathcal I}_\text{$4$D}(X'_1)-{\mathcal I}_\text{$4$D}(X'_2)={\mathcal I}_\text{$4$D}(X'_1\cup-X'_2)$ with $X'_{1,2}$ glued as FIG.~(\ref{glued}) through their common boundary $T^3$ along which the gauge fields are pasted by a gauge transformation, and the index for the closed manifold $X'_1\cup-X'_2$:
\begin{eqnarray}
\label{response}
&&{\mathcal I}_\text{$4$D}(X'_1\cup-X'_2)\nonumber\\
&=&2\left[-\frac{N}{48}\int_{X'_1\cup-X'_2}\frac{\text{tr}R\wedge R}{(2\pi)^2}-\frac{1}{2}\frac{\text{Tr}F\wedge F}{(2\pi)^2}\right]\nonumber\\
&=&-\frac{2}{32\pi^2}\int_{X'_1\cup-X'_2}d^4x\epsilon^{ijkl}\text{Tr}(F_{ij}F_{kl})\mod4N\nonumber\\
&=&1\mod4N,
\end{eqnarray}
with the gravitational contribution on general closed orientable manifolds valued in $2\times2\times N\mathbb{Z}=4N\mathbb{Z}$ thereby irrelevant to $Z_\pi$, and a unit $U(N)$ instanton inside $D_{X_1'}\cup-D_{X_2'}$ as shown in FIG.~(\ref{glued}).
Equation~(\ref{phase_ambiguity}) also means that the SPT bulk attached is nontrivial and the low-energy field theory (\ref{critical}) cannot be regulated by local regulators respecting onsite $U(N)$, $T_{1,2}$ and $\mathbb{Z}_2^T$ without the bulk attachment (\ref{bulk_attachment}).
}

The anomaly we have obtained is fully a mixed type since it will be trivialized once we discard any of the required symmetries.
This property is consistent with the ingappability on the lattice level, e.g. the magnetic-translation and $U(N)$ symmetric gapping term but time-reversal breaking $\Delta H$ in Eq.~(\ref{dH}) trivially gaps the system.
Thus we expect such a non-abelian symmetry anomaly is the desired LSM type. Moreover, two layers (or copies) of lattice systems can be trivially gapped respecting all the required symmetries since we can always set opposite chemical-potential term, e.g. $\sum_f\sum_{Z=1,2}(-1)^Z\sum_{\vec{r}}\delta\mu c^{\dagger Z}_{\vec{r},f}c^{Z}_{\vec{r},f}$ with $Z$ labelling the layers so that one of two is fully-filled while the other empty, without breaking (total) filling fraction and symmetries. This aspect is also implied by and consistent with our $\mathbb{Z}_2$ anomaly classification. In addition, there is no purely $U(N)\rtimes\mathbb{Z}_2^T$ anomaly, which implies that without $T_{1,2}$ we can trivially gap the system, consistent with the lattice situation as well.
Furthermore, we do not use the emergent properties of $T_{1,2}$ at low-energy limit, such as their finite cyclicality, so the corresponding emergent anomalies, if any, are not included.
Instead, we only make use of the gauge-invariant nontrivial commutator $T_1T_2T_1^{-1}T_2^{-1}=-1$ already held at the lattice level.
{Additionally, due to the time-reversal symmetry, we should also consider non-orientable manifold in Euclidean signature twisted by $CR_\tau$ (c.f. Eq.~(\ref{wick_rotation})). However, the anomaly class detected in this way is still $\mathbb{Z}_2$ identical to the $\mathbb{Z}_2$-class in our orientable cases which implies the absence of pure time-reversal anomaly~(See e.g. Sec.~IV.~G. of \cite{Witten:2016aa}).}
Combining these observations above, this mixed anomaly in field theory suggests the ingappability in the presence of the time reversal symmetry together with the flavor and magnetic translation symmetries.
This will be further confirmed in the next subsection, by defining a corresponding quantity on the lattice.

\subsection{Lattice-realization of $\mathbb{Z}_2$-classifying anomaly}

{
$Z_\pi=-1$ defined for the effective field theory in the previous subsection
is a topological invariant, e.g. it cannot be changed along renormalization flow (RG) from the critical point as long as the interaction perturbing the system respects all the required symmetries.
Here we look for a lattice quantity $Z'_\pi$ whose low-energy and continuum limit is $Z_\pi$. The 't Hooft anomaly matching then suggests it would be a robust topological invariant which can be evaluated exactly by its infrared counterpart $Z_\pi$.}

{
Let us assume that the system is trivially gapped under all the symmetries.
It will lead to a contradiction, as we will demonstrate below.}
To do so, we first impose the following TLBC:
\begin{eqnarray}
\label{xtlbc}
\left\{\begin{array}{l}c_{x+L_x,y}=c_{x,y+1}, \\
c_{x,y+L_y}=c_{x,y}. \end{array}\right.
\end{eqnarray}
The lattice translation $T_2$ is well-defined, but the magnetic translation $T_1$, {which is defined on the lattice before modded by the relation (\ref{xtlbc}), is incompatible with (\ref{xtlbc})} since it is $y$-dependent and $y$ coordinate has a freedom to be adjusted by Eq.~(\ref{xtlbc}).
{A sensible form of $T_1$ in the modded space can be obtained by fixing the assignment of lattice coordinates, e.g. $c_{x,0}$ with $x=1,\cdots,L_xL_y$ similar to Eq.~(\ref{coordinate_fix}). By Eq.~(\ref{gauge_fix}), $T_1$ reduces to $T'_1$ which is the lattice translation.}
The existence of $T_1$ on the lattice with periodic boundary conditions implies that such a symmetry has a different form in our tilted case especially in the low-energy sense.
{It follows straightforwardly from the assumption of the gapped unique ground state that the ground state should go back to itself after an adiabatic insertion $(T\rightarrow+\infty)$ of $\pi$ flux into the hole rounded by $L_x$: }
\begin{eqnarray}
A_x(t,\vec{r})=\frac{\pi}{L_x}\frac{t}{T}t_0
\end{eqnarray}
followed by the $T'_1$ transformation:
\begin{eqnarray}
T'_1c_{x,y}{T'_1}^{-1}=c_{x+1,y}.
\end{eqnarray}
It is due to that the following large gauge transformation $V$:
\begin{eqnarray}
V'c_{x,y}{V'}^{-1}=\exp\left( i \frac{\pi}{L_x}x\right)\exp( i \pi y)c_{x,y},
\end{eqnarray}
with $(x,y)\in[1,L_x]\times[1,L_y]$ is well-defined since the original magnetic translation $T_1$ requires $L_y\in2\mathbb{Z}$ as mentioned before, and it is able to restore the initial Hamiltonian after the preceding $\pi$-flux insertion and $T_1'$.

Let us consider the following quantity: ($\mathscr{T}$ is time-ordering)
\begin{eqnarray}
&&z^{\mbox{\scriptsize latt}}(\theta)\nonumber\\
&\equiv&\text{Tr}_\text{G.S.}\left\{\hat{T}'_1V'\mathscr{T}\exp\left\{- i \int H[A_x(t),A_y^\theta]dt\right\}\right\}\nonumber\\
&=&\left\langle\!\!\text{G.S.}\!\left|\hat{T}'_1V'\mathscr{T}\exp\left\{- i \int H[A_x(t),A_y^\theta]dt\right\}\right|\!\text{G.S.}\!\!\right\rangle, \nonumber
\end{eqnarray}
where $H$ is the time-dependent lattice Hamiltonian and $|\text{G.S.}\rangle$ is the certain presumed unique gapped ground state.
We also set an artificial flat $U(1)_Q$ gauge field in $y$ direction by $A_y=\theta(t_0-t_{N^2-1})/(NL_y)$ where our case corresponds to $\theta=0$.
Then $z^\text{latt}(\theta)$ satisfies:
\begin{eqnarray}
z^{\mbox{\scriptsize latt}}(\theta)=z^{\mbox{\scriptsize latt}}(\theta+2\pi),
\end{eqnarray}
due to the large gauge transformation mentioned before.

Let us imagine a series of model with $\theta$ changing from $0$ to $2\pi$
and consider
\begin{eqnarray}
\label{lattice_Z}
Z_\pi^\text{latt}&\equiv&\frac{z^{\mbox{\scriptsize latt}}[\theta=2\pi]}{z^{\mbox{\scriptsize latt}}[\theta=0]}.
\end{eqnarray}

Before evaluating this ratio of partition functions, we first notice that $T_1'V'$ acting on $c_{x,y}$ with $(x,y)\in[1,L_x]\times[1,L_y]$:
\begin{eqnarray}
&&T_1'V'c_{x,y}(T_1'V')^{-1}\nonumber\\
&=&\exp\left( i \frac{\pi}{L_x}x\right)\exp( i \pi y)c_{x+1,y}\nonumber\\
&=&T_1Vc_{x,y}(T_1V)^{-1},
\end{eqnarray}
which has exactly the same effect as the $T_1$ on periodic lattice followed by a ``half-large'' gauge transformation $Vc_{x,y}V^{-1}\equiv\exp( i \pi x/L_x)$ with $(x,y)\in[1,L_x]\times[1,L_y]$, which is well-defined according to the discussion below Eq.~(\ref{tlbc_tau}).

Therefore, the continuum limit of $z'$ is
\begin{eqnarray}
z^{\mbox{\scriptsize latt}}&=&\text{Tr}_\text{G.S.}\left\{\hat{T}'_1V'\mathscr{T}\exp\left\{-\int H[A_x(\tau),A_y^\theta]d\tau\right\}\right\}\nonumber\\
&\rightarrow&\text{Tr}_\text{G.S.}\left\{\hat{T}_1V\mathscr{T}\exp\left\{-\int H[A_x(\tau),A_y^\theta]d\tau\right\}\right\}, \nonumber
\end{eqnarray}
which precisely reproduces the form of $z$ in Eq.~(\ref{z}) with the boundary condition as Eq.~(\ref{tlbc_tau}).
Similarly as the $U(1)_Q$ case, the role played by $V$ in the path integral of $z$ is implicitly an assignment of a transition function between the last two time slices preceding the $\hat{T}_1$ gluing.
Thus, we can make use of 't Hooft anomaly matching to evaluate $Z_\pi^\text{latt}$ by $Z_\pi$:
\begin{eqnarray}
Z^\text{latt}_\pi=Z_\pi=-1.
\end{eqnarray}
Therefore, we have
\begin{eqnarray}
z^\text{latt}(\theta=0)=z^\text{latt}(\theta=2\pi)=-z^\text{latt}(\theta=0),
\end{eqnarray}
which is impossible unless
\begin{eqnarray}
z^\text{latt}(\theta=0)=0.
\end{eqnarray}
It implies the following two equal-energy states are actually orthogonal:
\begin{eqnarray}
\left.\left.\hat{T}'_1V'\mathscr{T}\exp\left\{- i \int_0^TH[A_x(t),A_y=0]dt\right\}\right|\!\text{G.S.}\!\!\right\rangle\!\perp\!|\text{G.S.}\rangle, \nonumber
\end{eqnarray}
thereby contradicting the assumption of the unique $|\text{G.S.}\rangle$. It means the ground-state degeneracy must be nontrivial and we arrive at an LSM-type ingappabillity.

\section{TLBC applied to many-body Chern numbers}
\label{IQHE}
In this Section, we will apply TLBC to obtain a nontrivial constraint on integer quantum Hall conductances. The well-known Thouless-Kohmoto-Nightingale-Nijs
formula~\cite{Thouless:1982aa,Kohmoto1985}
identifies the Hall conductance of a band insulator
with the Chern number defined by the Berry connection of
eigenstates in the Brillouin zone.
It was then generalized to ``many-body Chern number'' for
more general interacting systems~\cite{NiuThoulessWu1985}.
It should be noted that, by definition, evaluation of
the Chern number generally requires an integral over the
entire Brillouin zone or the entire parameter space.
{Such integrations can be time-consuming
even for free electrons (band insulator) with generic band structures,
{especially when Chern numbers $n_\text{ch}$ are large,} since the critical mesh size of the discretized Brillouin zone is of order ${\mathcal O}(\sqrt{n_\text{ch}})$~\cite{Fukui:2005aa}.}
{Evaluation of many-body Chern numbers of interacting
systems can be even more difficult
(however, see also Ref.~\cite{Kudo-MBChern2019}).
}

 {
On the other hand, since the low-energy physics of lattice models
can be often described by field theory, such as Dirac fermions,
it is natural to attempt to determine the Hall conductance (Chern number)
by the low-energy effective field theory.
If this works, the Chern number appears to be determined with
only the information in the low-energy limit.
However, the most naive expectation fails
in general~\cite{Oshikawa:1994aa}.
This is not surprising, considering the global nature of the Chern number.
In terms of field theory, the discrepancy can be attributed
to so-called ``spectator'' fermions which may exist at higher
energies and are invisible in low-energy
physics~\cite{Hatsugai:1996aa,Lee:1998aa,Watanabe:2010aa,Bernevig:2013aa}.
}

 {
Nevertheless,
anomaly, which is a central concept utilized in the present paper,
is a essential property of field theory which is robust even
at higher energies.
We can therefore derive a nontrivial constraint on the Hall conductance
from field theory, as we demonstrate below.
In other words, the anomaly controls possible behaviors of the spectator
fermions.
}

 {
Let us consider the half-filled $\pi$-flux system above. Since the system can be gapped if we explicitly break the time-reversal symmetry, e.g. by $\Delta H$ in Eq.~(\ref{dH}), it is gappable with a unique ground state in the presence of $U(N)$ and magnetic translations. In the following discussion, we will investigate the constraint on the many-body Chern number
{in such a gapped phase}
when $U(N)$ and magnetic translations are respected by the half-filled spinless lattice Hamiltonians.
}

 {
To derive the constraint, however, it is convenient to start from
the gapless system which is obtained by imposing the time-reversal
symmetry.
By TLBC in Eq.~(\ref{tlbc}) together with its lattice realization in Eq.~(\ref{xtlbc}), we have obtained the partition function Eq.~(\ref{bulk_edge_correspondence}) for the critical Hamiltonian $H_\pi$ in Eq.~(\ref{eq.Hpi}).
It can be rewritten as
\begin{eqnarray}
\label{bulk_edge_correspondence_1}
Z_\text{$3$D}(X')&=&Z_\text{$3$D:P-V}\exp\left[i\pi\int_{X':\partial X'=T^3}\frac{\text{Tr}F\wedge F}{(2\pi)^2}\right], \nonumber\\
\end{eqnarray}
where $Z_\text{$3$D:P-V}\equiv|Z_\text{$3$D}|\exp(-i \pi{\eta_\text{$3$D}}/{2})$ is regularized by the Pauli-Villars regulator defined before, which is automatically gauge invariant, and $T^3$ is the $(2+1)$-dimensional spacetime where our square lattice is defined. At the first sight, the partition function depends on the choice of the $4$-dimensional spacetime $X'$. However, as we will see, the physical observables, such as Hall conductances, when we gap it, is independent of this dimension extension. In addition, the $X'$-bulk regulator is the only sensible way to regularize the partition function up to a non-universal part as summarized by \cite{Witten:2019aa}.
}

 {
By the discussion following (\ref{bulk_edge_correspondence}), the partition function (\ref{bulk_edge_correspondence_1}) is, up to a non-universal factor, equal to the partition function of a $(3+1)$-dimensional Dirac fermion in $\mathbb{R}^4\supset X'$ whose mass is opposite of the P-V regulator within $X'$ and is the same as the P-V regulator mass in $(\mathbb{R}^4-X')$, which can be identified as the vacuum [c.f. (\ref{spt_bulk}) and (\ref{spt_vac})]~\cite{Seiberg:2016aa,Fukaya:2017aa,Witten:2019aa} as in FIG.~(\ref{interface}).
\begin{figure}
\begin{center}
\includegraphics[width=6cm,pagebox=cropbox,clip]{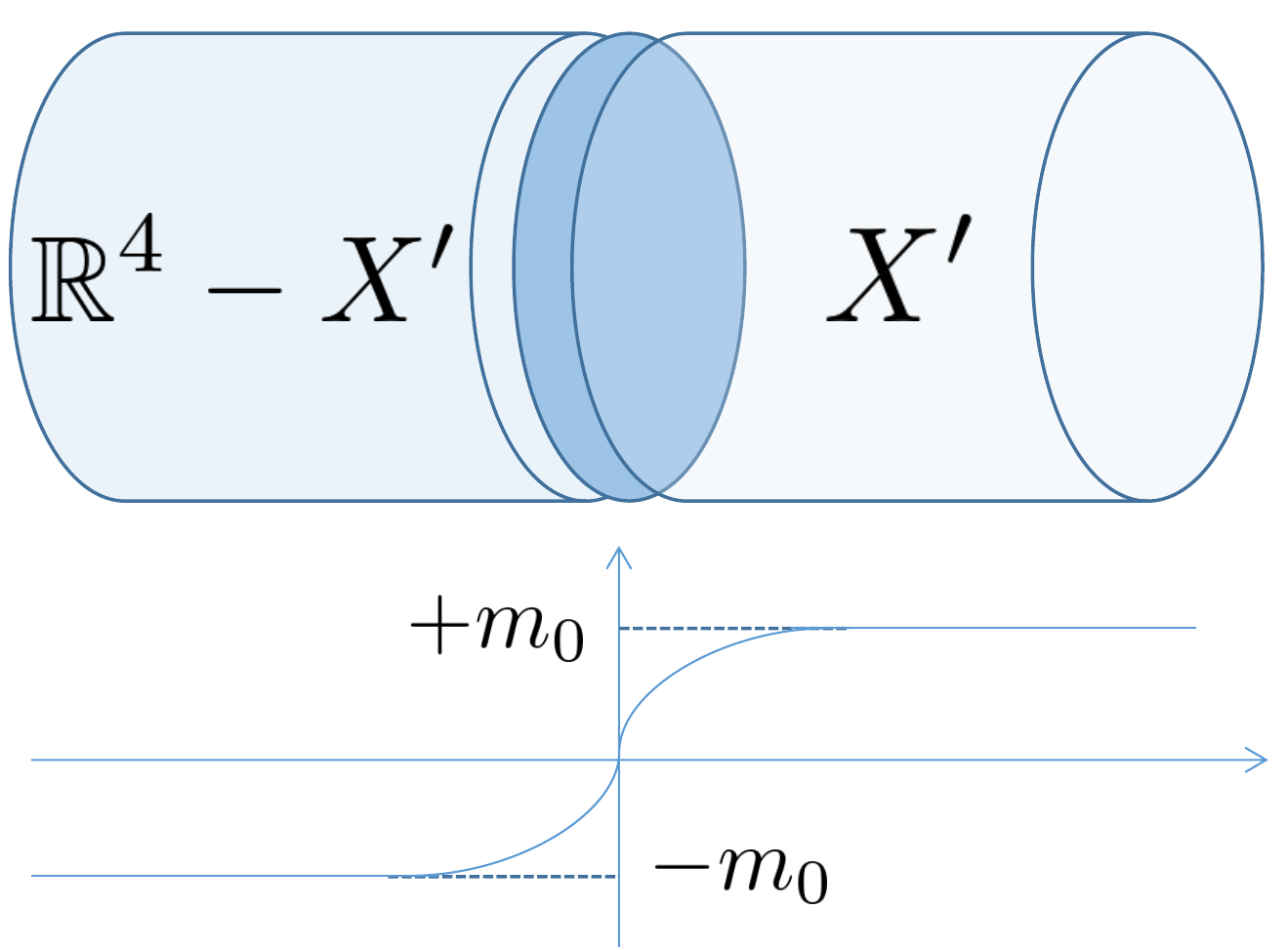}
\caption{The bulk-interface-bulk equivalence of the partition function (\ref{bulk_edge_correspondence_1}) where the P-V regulator mass is $-m_0$ in (\ref{spt_bulk}) and (\ref{spt_vac}). }
\label{interface}
\end{center}
\end{figure}
The $(3+1)$-dimensional Dirac mass inevitably vanishes on the interface $T^3=\partial X'=-\partial\left(\mathbb{R}^4-X'\right)$, along which the $(2+1)$-dimensional gapless mode flows, {because the masses have different signs on the two sides of the interface and the mass is restricted to be real throughout the whole bulk by the time-reversal symmetry}.
Then we introduce a general {local} interaction, not necessarily perturbative, to the Hamiltonian $H_\pi$ in Eq.~(\ref{eq.Hpi}), and the interaction gaps the system with a unique ground state, breaking time-reversal symmetry while respecting $U(N)$ and magnetic translations.
In the equivalent $(3+1)$-dimensional bulk-interface-bulk picture described above, the locality of this gapping interaction in $(2+1)$ dimensions is translated as follows: the interaction only takes place locally around the interface $\partial X'=T^3$ (since it is local on $\partial X'$) and thus the time reversal is still preserved deeply in the $X'$ and the vacuum $(\mathbb{R}^4-X')$ on its two sides.
Then the partition function for this gapped system takes the following general universal form by the interface interpretation above (c.f. Appendix):
\begin{eqnarray}
\label{gapped_1}
Z_\text{$3$D-gapped}(X')=\exp\left[i\int_{\mathbb{R}^4}\theta(x)\frac{\text{Tr}F\wedge F}{(2\pi)^2}\right],
\end{eqnarray}
with
\begin{eqnarray}
\theta(x)=\left\{\begin{array}{ll}-2n\pi,&x\text{ deep in the vacuum};\\\pi,&x\text{ deep in }X', \end{array}\right.
\end{eqnarray}
where the universal quantity $n\in\mathbb{Z}$ since the theta-angle of the vacuum is $0\mod2\pi$ and the varying $\theta(x)$ only breaks time reversal near $\partial X'=T^3$~\footnote{Here we fix the theta-angle deep in $X'$ but relax that in the vacuum without loss of generality since only their difference is physical.}.
(A non-local gapping interaction is able to change the $\theta$-angle arbitrarily deeply inside $X'$ or the vacuum.)
Thus, by an integration by part for the exponential in Eq.~(\ref{gapped_1}) and taking a $U(1)_Q$ connection $A=A_{u(1)}\mathbb{I}_{N\times N}$, we obtain
\begin{eqnarray}
\label{response_gapped}
Z_\text{$3$D-gapped}(X')=\exp\left[i\int_{T^3}\frac{(2n+1)N}{4\pi}A_{u(1)}\wedge dA_{u(1)}\right], \nonumber\\
\end{eqnarray}
which is explicitly a pure $(2+1)$-dimensional partition function on $T^3$ independent of the ``bulk'' $X'$, as expected since our system is purely supported by a $2$-dimensional spatial lattice.
{The physical meaning of the winding number ``$n$'' of $\theta(x)$ is the stack of $n$ layers of gapped $U(N)$ and magnetic-translation symmetric systems, e.g. $n$ layers of $2$ copies of our boundary theories (\ref{critical}) with mass terms, which are able to non-anomalously respect the time reversal in addition to $U(N)$ and $T_{1,2}$ at the low energy in their own dimensions. }

We can directly, by the level of the Chern-Simons term in Eq.~(\ref{response_gapped}), identify the Hall conductance or in terms of many-body Chern number as:
\begin{eqnarray}
\label{hall}
\sigma_\text{H}=N\mod2N.
\end{eqnarray}
{
The permitted Hall conductances by the constraint above are symmetric about the zero value, which is not a coincidence. Let us assume that, preserving the $U(N)$ and magnetic translations, we have realized an integer quantum Hall state with $\sigma^{(0)}_\text{H}$. Then we do a time-reversal transformation $\Theta_0$ on this state (or its parent Hamiltonian) and the resultant state has the Hall conductance $-\sigma^{(0)}_\text{H}$, which still respects the original $U(N)$ since, the group elements of $\Theta_0^{-1}U(N)\Theta_0$ are the same as $U(N)$, which can be seen in a basis where all the $U(N)$ generators are purely real or imaginary~\footnote{This can be always done for general $N$'s since the generators are Hermitian.}. This state is also symmetric under the original magnetic translations since the gauge-invariant commutator (\ref{commutator_inv}) is time-reversal invariant (namely, $\pi$-flux is time-reversal symmetric). Thus, the permitted Hall conductances must be symmetric about the zero. For general flux cases out of the scope, the above argument does not work, since if we want to map the magnetic translations back to its initial form, we need to further do a spatial reflection which, together with the time reversal, trivially transforms $\sigma_\text{H}^{(0)}$ back to $\sigma_\text{H}^{(0)}$.
}

{This generalizes the constraint for $N=1$
obtained from the lattice argument~\cite{AvronYaffe1986,KolRead1993,Lu:2017aa,Matsugatani:2018}.
It is notable that this generalized
constraint can be derived within the field theory.}
We can see that the contribution from the spectator fermion is $0\mod2N$ which is controllable in the presence of $U(N)$ and magnetic translations.

{In a short summary, TLBC is useful in the bulk SPT identification of the gapless point Eq.~(\ref{critical}) since it constitutes a gauge bundle in Eq.~(\ref{tlbc}) to detect the anomaly Eq.~(\ref{phase_ambiguity}). This bulk SPT correspondence is helpful since it provides us a visualizable way to treat our $(2+1)$-dimensional system as a domain wall in FIG.~(\ref{interface}) and, furthermore, to restrict the integer Hall conductances. }
}

 {
We can further argue that such a {constraint on many-body Chern numbers} is part of a more general framework.
An observation of Eq.~(\ref{gapped_1}) implies that we can imagine a general system as a $\mathbb{Z}_2^T$-broken surface of an SPT bulk protected by $(G\times\mathbb{Z}^2)\rtimes\mathbb{Z}_2^T$ where $G$ is an onsite symmetry and $\mathbb{Z}^2$ can be translations or magnetic translations. Then the bulk regularization can induce a natural constraint on the $G$-response, e.g. Hall conductances. The surface theory is trivial in the sense that the bulk is trivial if $\mathbb{Z}_2^T$ is broken, while the nontrivial property, e.g. constraints of Hall conductances, results from that the surface theory can be attached to nontrivial SPT bulk where $\mathbb{Z}_2^T$ is preserved deeply in bulk. In addition, the method is also directly applicable to arbitrary even spatial dimensions where Chern-Simons actions can be defined.
}

\section{Conclusions and Discussions}
In this work, we introduced a generalized boundary condition TLBC for discussion of the LSM theorem and related problems.
Under the TLBC, each lattice site can be reached by repeating the lattice translation in a particular direction.
As an advantage of the TLBC, the LSM theorem in two or more dimensions can be derived by the flux insertion argument without an artificial restriction on the system sizes, which renders the thermodynamic limit more natural.
Moreover, with the TLBC, the LSM theorem in arbitrary dimensions is related to the global chiral anomaly of Dirac fermion in $(1+1)$ dimensions in a unified manner.
This is also extended to the LSM theorem with the larger $SU(N)$ symmetry, and to the similar ingappability constraint under the time reversal and the magnetic translation symmetries with the on-site $U(N)$ symmetry.
 {Furthermore, we also utilized the TLBC to derive a nontrivial constraint
on many-body Chern numbers from field theory, which generalizes
the known result from the lattice.
It shows the power of anomaly which universally dictates higher energies,
within the universal field theory formulation and independently of the
regularization/realization.
Although similar phenomena have been
proposed on the surface of topological insulators~\cite{Qi:2008aa}, our
result is the first to relate the Hall conductance of a pure
$2$-dimensional lattice system to SPT phases.
}

Despite of the usefulness of TLBC as we have demonstrated, it cannot be
applied to exploit consequences of spatial reflection symmetries, which
are inconsistent with the TLBC as introduced in this paper.
Nevertheless, we believe that the idea of modifying the boundary
condition from the standard PBC in the LSM-type argument,
and combining it with anomaly in field theory, is opening up a new
direction in quantum many-body theory.
 {In fact, after the appearance of the early version of the present
paper in arXiv, Furuya and Horinouchi~\cite{Furuya:2019aa}
generalized our work by introducing a twisted boundary condition
and deriving a nontrivial constraint for systems on the checkerboard lattice.
We hope that more developments will follow in the future.
}


\acknowledgements
Both authors thank Yohei Fuji, Shunsuke Furuya, Hans Hansson, Chang-Tse Hsieh, Yuan-Ming Lu, Ying Ran, Shinsei Ryu, Xiaolin Sun, Yuji Tachikawa, Yasuhiro Tada, Cenke Xu, and Xu Yang for useful discussions.
Y.~Y. was supported by JSPS fellowship.
This work was supported in part by MEXT/JSPS KAKENHI Grant Nos. JP19J13783 (Y.~Y.), JP17H06462 (M.~O.) and JP19H01808 (M.~O.).
A part of the present work was performed at Kavli Institute for Theoretical Physics, University of California at Santa Barbara, supported by
US National Science Foundation Grant No. NSF PHY-1748958.

\appendix*
\section{Derivations of (\ref{gapped_1}), (\ref{response_gapped}) and (\ref{hall})}

In this part, we will give a rigorous proof of (\ref{gapped_1}) and (\ref{response_gapped}).

Due to the discussion around Eq.~(\ref{bulk_edge_correspondence}), the low-energy effective theory can be thought as the interface theory between the vacuum and a bulk in the Euclidean signature as
\begin{eqnarray}
\label{spt_bulk}
S_\text{bulk}=\sum_{\alpha,f}\int_{X'} i\bar{\psi}_{\alpha,f}(-i\mathscr{D}_4-m_0)\psi_{\alpha,f}+S_\text{reg},
\end{eqnarray}
where $S_\text{reg}\equiv\sum_\alpha\int_{\mathbb{R}^4} i\bar{\chi}_{\alpha,f}(-i\mathscr{D}_4+m_0)\chi_{\alpha,f}$ is the bosonically statistical spinor regulator and $\mathscr{D}_4=\sum_{i=0}^3\tilde{\Gamma}^i(\partial_i-\mathbb{I}A_i)$ is the $4$-D Dirac operator where $\{\tilde{\Gamma}^i,\tilde{\Gamma}^j\}=-2\delta^{ij}$, $i,j\in\{0,1,2,3\}$. The vacuum action is simply
\begin{eqnarray}
\label{spt_vac}
S_\text{vac}=\sum_{\alpha,f}\int_{\mathbb{R}^4-X'} i\bar{\psi}_{\alpha,f}(-i\mathscr{D}_4+m_0)\psi_{\alpha,f}+S_\text{reg}. \nonumber\\
\end{eqnarray}
The symmetries $U(N)$ and $T_{1,2}$ all extends in a natural way to both bulks. Gapping our $(2+1)$-dimensional system by a $U(N)$ and $T_{1,2}$ gapping term can be seen as physically equivalent to gapping the interface between the artificial bulks $S_\text{bulk}$ and $S_\text{vac}$ by a $U(N)$ and $T_{1,2}$ respecting boundary interaction within the interface. Therefore, we can resolve the former question in the approach to the later one.

Since $X'$ is embedded in $\mathbb{R}^4$ in a natural way, we take a local geometry near $T^3=\partial X'$ as $T^3\times\mathbb{R}$ where $\mathbb{R}:x^3\in(-\infty,+\infty)$.
We introduce a spatial dependence $U(N)$-preserving mass term $m^{\alpha\beta}(x^3)$ to gap the boundary states where $\alpha$ and $\beta$ are valley indices, namely the whole material to be gapped throughout $x_3\in(-\infty,\infty)$:
\begin{eqnarray}
S_\text{full}&=&\sum_{\alpha,\beta}\int_{\mathbb{R}^4} i\bar{\psi}_{\alpha,f}\left[-i\mathscr{D}_4\delta^{\alpha\beta}-m^{\alpha,f;\beta,f'}(x^3)\right]\psi_{\beta,f'}\nonumber\\
&&+S_\text{reg},
\end{eqnarray}
where $m^{\alpha,f;\beta,f'}(x^3)$ interpolates the nontrivial bulk $S_\text{bulk}$ asymptotical at $x^3\rightarrow+\infty$ and the vacuum at $x^3\rightarrow-\infty$:
\begin{eqnarray}
m^{\alpha,f;\beta,f'}(x^3)=\left\{\begin{array}{ll}m_0\delta^{\alpha,\beta}\delta^{f,f'},&x^3\rightarrow+\infty;\\-m_0\delta^{\alpha,\beta}\delta^{f,f'},&x^3\rightarrow-\infty. \end{array}\right.
\end{eqnarray}
Moreover, the gapping is required to respect $U(N)$ and $T_{1,2}$ proportional to $\tau_{3,1}$, and then
\begin{eqnarray}
m^{\alpha,f;\beta,f'}(x^3)\propto\delta^{\alpha,\beta}\delta^{f,f'}
\end{eqnarray}
since any matrix in $SL(2,\mathbb{C})$ commuting with $\tau_{1,3}$ must be proportional to identity.
Combining this result and that a general non-vanishing mass term can only take a Dirac-mass or a chiral-mass form, then the interpolating action is
\begin{eqnarray}
S_\text{full}&=&\sum_{\alpha,f}\int_{\mathbb{R}^4} i\bar{\psi}_{\alpha,f}\left\{-i\mathscr{D}_4+m_0\exp\left[i\tilde{\Gamma}^5\theta(x^3)\right]\right\}\psi_{\alpha,f}\nonumber\\
&&+S_\text{reg},
\end{eqnarray}
where we have, without loss of generality, normalized the mass amplitude to be a constant $m_0$ which can be done canonically, since the mass gap never closes for any $x^3$, and $\tilde{\Gamma}^5\equiv-\tilde{\Gamma}^0\tilde{\Gamma}^1\tilde{\Gamma}^2\tilde{\Gamma}^3$ is Hermitian and the continuous function $\theta(x^3)$ satisfies
\begin{eqnarray}
\label{theta_winding}
\theta(x^3)=\left\{\begin{array}{ll}\pi\mod2\pi,&x^3\rightarrow+\infty;\\0\mod2\pi,&x^3\rightarrow-\infty. \end{array}\right.
\end{eqnarray}
To evaluate the partition function of $S_\text{full}$, we do the following $x^3$-dependent chiral rotation
\begin{eqnarray}
\psi_{\alpha,f}&\rightarrow&\exp\left[i\tilde{\Gamma}^5\theta(x^3)/2\right]\psi_{\alpha,f};\nonumber\\
\bar{\psi}_{\alpha,f}&\rightarrow&\bar{\psi}_{\alpha,f}\exp\left[i\tilde{\Gamma}^5\theta(x^3)/2\right]
\end{eqnarray}
to eliminate the chiral phase of the mass term. However, it is inevitable to introduce a theta term due to $4$-D global chiral anomaly although the matter partition function is regularized to be unity by $S_\text{reg}$. Therefore, after a $U(1)_Q$ connection $A\equiv At_0$ is taken,
\begin{eqnarray}
Z_\text{full}&=&\exp\left[i\int_{\mathbb{R}^4}\theta(x^3)\frac{\text{Tr}F\wedge F}{(2\pi)^2}\right]\nonumber\\
&=&\exp\left\{\frac{iN}{(2\pi)^2}\int_{\mathbb{R}^4}\left\{d\left[\theta\wedge d(A\wedge dA)-d\theta\wedge A\wedge dA\right]\right\}\right\}\nonumber\\
&=&\exp\left[i\int_{T^3}\left(\frac{\int d\theta}{\pi}\right)\frac{N}{4\pi}A\wedge dA\right],
\end{eqnarray}
where we have made use of Bianchi identity $d(dA)=0$ for $U(1)_Q$ connections and the assumption that $A_\mu$ is nearly $x^3$-independent within the interface of two bulks.

The response exactly implies that, at the interface, namely our interested $(2+1)$-dimensional system on $T^3$,
\begin{eqnarray}
\sigma_\text{H}&=&{N\int dx^3\partial_{x^3}\theta}/{\pi}\nonumber\\
&=&N\mod2N,
\end{eqnarray}
where we have used the winding property of $\theta(x^3)$ in Eq.~(\ref{theta_winding}). Thus we reach the final conclusion of the constraint on the Hall conductance of the uniquely gapped $U(N)$ and $T_{1,2}$ respecting system: $\sigma_\text{H}=N\mod2N$.
The role played by $T_{1,2}$ is also obvious from the derivation above, e.g. without $T_1$ or $T_2$, we can gap the boundary, respectively, by a mass as $m_0\exp\left[\tau_3\otimes i\tilde{\Gamma}^5\theta(x^3)\right]\delta^{f,f'}$ or $m_0\exp\left[\tau_1\otimes i\tilde{\Gamma}^5\theta(x^3)\right]\delta^{f,f'}$, resulting an unrestricted $\sigma_\text{H}$.

\sloppy
%

\end{document}